\documentclass[11pt]{article}
\usepackage{latexsym}
\usepackage{boxedminipage}
\usepackage{graphicx}
\usepackage[authoryear]{natbib}
\usepackage{amsmath,amsthm,amssymb}
\usepackage{tikz}
\usepackage{enumitem}

\usetikzlibrary{arrows,chains,matrix,positioning,scopes,shapes.multipart}

\DeclareMathOperator*{\argmax}{arg\,max}
\makeatletter
\tikzset{join/.code=\tikzset{after node path={%
\ifx\tikzchainprevious\pgfutil@empty\else(\tikzchainprevious)%
edge[every join]#1(\tikzchaincurrent)\fi}}}
\makeatother
\tikzset{>=stealth',every on chain/.append style={join},
         every join/.style={->}}
\tikzstyle{labeled}=[execute at begin node=$\scriptstyle,
   execute at end node=$]

\def\IR{\rm I \kern-0.20em R}

\newcommand{\be}{\begin{eqnarray}}
\newcommand{\ee}{\end{eqnarray}}
\newcommand{\ben}{\begin{eqnarray*}}
\newcommand{\een}{\end{eqnarray*}}
\newcommand{\utwi}[1]{\mbox{\boldmath $ #1$}}

\newcommand{\bX}{{\utwi{X}}}
\newcommand{\bY}{{\utwi{Y}}}

\newcommand{\cI}{{\cal I}}

\bibpunct{(}{)}{;}{a}{,}{,}

\begin{document}
\setcounter{page}{1}

\title{\bf Resampling Strategy in Sequential Monte Carlo for Constrained Sampling Problems
\footnote{
Rong Chen's research was supported in part by National Science Foundation
grants DMS-1503409, DMS-1737857 and IIS-1741390.
Corresponding author: Rong Chen,
Department of Statistics, Rutgers University, Piscataway, NJ 08854,
USA. Email: rongchen@stat.rutgers.edu. }}
\author{Chencheng Cai, Rong Chen and Ming Lin
\\ \vspace{0.2cm} {Rutgers University and Xiamen University}\\}
\date{}
\maketitle


\begin{abstract}
Sequential Monte Carlo (SMC) methods are a class of Monte Carlo methods that are used to obtain random samples of a high dimensional random variable in a sequential fashion. Many problems encountered in applications often involve different types of constraints.
These constraints can make the problem much more challenging. In this paper, we formulate a general framework of using SMC for constrained sampling problems based on forward and backward pilot resampling strategies.
We review some existing methods under the framework and develop several new algorithms.
It is noted that all information observed or imposed on the underlying system can be viewed as constraints.
Hence the approach outlined in this paper can be useful in many applications.

\end{abstract}
\noindent {\bf Keywords:} Backward sampling, Constrained sampling,
Pilot, Priority score, Resampling, Sequential Monte Carlo

\newpage

\section{Introduction}
\par
Stochastic dynamic systems are used to model the dynamic behavior of random variables in a wide range of applications in physics, finance, engineering and other fields.
One of the important problems of studying complex dynamic systems is to sample paths following the underlying stochastic process.
Such paths can be used for statistical inferences under the Monte Carlo framework. In practice, a stochastic system often comes with observable information, including direct/indirect measurements, external constraints and others.
For example, in a state-space model, noisy measurements of the underlying latent states are observed. In a diffusion bridge sampling problem, the start and end points of the diffusion process are fixed. In this paper, we take the view that the underlying system is given and all available information is treated as imposed constraints.

The Sequential Monte Carlo (SMC) methods are a class of sampling methods that utilize the sequential nature of the underlying process. It has a wide range of applications \citep{kong1994sequential, avitzour1995stochastic,liu1995blind,kitagawa1996monte, kim1998stochastic,pitt1999filtering, chen2000adaptive, doucet2001introduction, fong2002monte, godsill2004monte}.
The sequential importance sampling with resampling (SISR) scheme embedded in SMC enables sampling from complex target distributions \citep{gordon1993novel, kong1994sequential,liu1998sequential}.
However, the choice of proposal distribution and the choice of priority score for resampling in SISR are crucial to sample quality and inference efficiency. For example, in a diffusion bridge sampling problem, where the start and end points of a diffusion process are exactly enforced,
\cite{pedersen1995consistency} proposed to generate the samples through the underlying diffusion process without considering the endpoint constraint and then force the samples to connect with the fixed point at the end. It may not be efficient due to the large deviation of the end of the forward paths from the enforced end point.
\cite{durham2002numerical} proposed a method based on SMC with linear interpolation as the proposal distribution. It ignores the drift term of the underlying diffusion process and may not be efficient for non-linear processes. \cite{lin2010} generated bridge samples based on a backward pilot resampling strategy.
In their procedure, a pilot run is conducted backward from the fixed end point to determine the priority scores for resampling in a forward SISR procedure.
The backward pilot resampling strategy achieves good efficiency.
This approach improves the forward sampling by bringing future information and constraints for effective sampling, especially with minimum additional computational costs.

In this article, we extend the procedure of \cite{lin2010} to more general settings. Specifically, the problem of simulating a stochastic process under constraints is more formally stated in a general setting that contains many problems as its special cases, including the standard state space models.
The general setting also allows the discussion of a formal guidance for improving efficiency in developing SMC implementations for such problems.
Under the setting, we propose a general framework for constrained sampling problems with measure theoretic interpretations.
Resampling strategies based on forward pilots and backward pilots are developed under such a framework.
Several types of constraints are discussed along with their corresponding SMC implementations. Links to the existing procedures are also discussed.
The developed approaches are demonstrated with several examples.

The rest of this paper is organized as follows.
In Section 2, the constrained sampling problem is formally stated and a general framework of the constrained SMC (cSMC) method is proposed.
Section 3 introduces some special constrained problems with their corresponding implementations under the cSMC framework.
Section 4 presents several methods to estimate the priority scores used in the resampling step of cSMC.
Three examples are used to demonstrate performance of the proposed methods in Section 5. Section 6 concludes.

\section{Constrained Sampling Problems}

\subsection{Stochastic Dynamic System with Constraints}

The stochastic system we consider contains a sequence of unobservable random states $x_{0:T}=\{x_0,x_1,\cdots,x_T\}$, whose dynamics is governed by an initial state distribution $p(x_0)$ and a known forward propagation distribution $p(x_t\,|\, x_{0:t-1})$.
In addition, a compounded information/constraint set $\mathcal{I}_{0:T}$ imposed on the latent states is given, where $\mathcal{I}_{0:T}=\mathcal{I}_0\cap \mathcal{I}_1\cap \cdots \cap \mathcal{I}_T$ and $\mathcal{I}_t$ is the new available information at time $t$.
For example, if we have a noisy measurement  $y_t=g(x_t,\varepsilon_t)$ at time $t$, then $\mathcal I_t=\{y_t\}$. If $x_t$ is a fixed point via prior knowledge or design,  we have  $\mathcal I_t=\{x_t=c\}$.
When there is no additional information at time $t$, we define $\mathcal I_t=\{x_t\in\mathcal{X}\}$, a trivial constraint, where $\mathcal{X}$ is the support of the state variable.
We further assume that
\ben p( x_t, \mathcal{I}_t\,|\, x_{0:t-1},\mathcal{I}_{0:t-1})=p( x_t, \mathcal{I}_t\,|\, x_{0:t-1})\een for any $t$,
that is, the past constraints only affect the future through the past states.

We focus on simulating the full path of $x_{0:T}$ under the forward propagation distribution $p(x_t\,|\,x_{0:t-1})$ and the given constraint set $\mathcal{I}_{0:T}$.
The posterior joint distribution of $x_{0:T}$ can then be written as
 \ben 
p(x_{0:T}\,|\,\mathcal I_{0:T}) =
p(x_0\,|\,\mathcal I_{0:T})\prod_{t=1}^T p(x_t\,|\,x_{0:t-1}, \mathcal{I}_{0:T}).
 \een
It induces a sequence of marginal posterior probability measures $\mathbb P_0,\mathbb P_1,\dots,\mathbb P_T$ with densities
\be
 \mathbb P_t (x_{0:t}) = p(x_{0:t}\,|\,\mathcal I_{0:T}) = p(x_0\,|\,\mathcal{I}_{0:T})\prod_{s=1}^t
p(x_s\,|\,x_{0:s-1},\mathcal{I}_{0:T}) \label{eq:marginal-dist}
\ee
for $t = 0,1,\dots, T$.
Note that $\mathbb P_t(x_{0:t})$ is defined under the full information set $\mathcal I_{0:T}$.
The recursion relationship  $\mathbb P_t(x_{0:t}) =\mathbb P_{t-1}(x_{0:t-1})\cdot p(x_t\,|\,x_{0:t-1},\mathcal I_{0:T})$ reveals a way to update samples to $x_{0:t}=(x_{0:t-1},x_t)$ given $x_{0:t-1}$.
However, under this sequence of measures, the conditional distribution $p(x_t\,|\,x_{0:t-1},\mathcal I_{0:T})$ is usually difficult to sample from, since it involves the entire information set from $t=0$ to $T$.
Hence such a direct sequential sampling procedure cannot be practically applied under this setting.

\subsection{Constrained Sequential Monte Carlo}

For a given sequence of forward propagation probability measures $\mathbb Q_0,\mathbb Q_1,\cdots,\mathbb Q_T$ with densities $\mathbb Q_0(x_0)$, $\mathbb Q_1(x_{0:1})$, $\cdots$, $\mathbb Q_T(x_{0:T})$, the sequential Monte Carlo (SMC) approach \citep{kitagawa1996monte,liu1998sequential, doucet2001introduction} proposes to generate samples $x_0^{(i)},x_1^{(i)},\cdots$, $i=1,\cdots,n$, sequentially from a series of proposal conditional distributions $q(x_t\,|\, x_{0:t-1})$, $t=0,1,\cdots$, and update the corresponding importance weights by
\ben w_t^{(i)}&=&\frac{\mathbb Q_t(x_{0:t}^{(i)})}{q(x_0^{(i)})\prod_{s=1}^t q(x_s^{(i)}\,|\, x_{0:s-1}^{(i)})}=w_{t-1}^{(i)}u_t^{(i)},\een
where $w_0^{(i)}=\mathbb{Q}_0(x_0^{(i)})/q(x_0^{(i)})$ and
\ben u_t^{(i)}=\frac{\mathbb Q_t(x_{0:t}^{(i)})}{\mathbb{Q}_{t-1}(x_{0:t-1}^{(i)})q(x_t^{(i)}\,|\, x_{0:t-1}^{(i)})},\quad t=1,2,\cdots,\een
which is called the {\it incremental weight}.
Under the principle of importance sampling, when $\mathbb{Q}_s(x_{0:t})$ is absolutely continuous with respect to $q(x_0)\prod_{s=1}^t q(x_s\,|\, x_{0:s-1})$, the sample set $\{(x_{0:t}^{(i)},w_t^{(i)})\}_{i=1,\cdots,n}$ is properly weighted with respect to $\mathbb{Q}_t$ at each time $t$, that is,
\ben
\frac{\sum_{i=1}^n w_t^{(i)}h(x_{0:t}^{(i)})}{\sum_{i=1}^nw_t^{(i)}} \stackrel{a.s.}{\longrightarrow} \mathbb E_{\mathbb{Q}_t}\big[h(x_{0:t})\big]
\een
as $n\rightarrow \infty$ for any measurable function $h(\cdot)$ with finite expectation under $\mathbb{Q}_t$.
The choice of the proposal distribution $q(x_t\,|\, x_{0:t-1})$ has a direct impact on the efficiency.
\cite{kong1994sequential} and \cite{liu1998sequential} proposed to choose \ben q_t(x_t\,|\, x_{0:t-1})=  \mathbb{Q}_t(x_t\,|\, x_{0:t-1})\een to minimize the variance of the incremental weight $u_t$ conditional on $x_{0:t-1}$.
In this case, we have $u_t =\mathbb{Q}_t(x_{0:t-1})/\mathbb{Q}_{t-1}(x_{0:t-1})$.

To use the SMC approach, we note that if at the ending time $T$, the forward propagation measure $\mathbb Q_T$ agrees with the
posterior measure $\mathbb P_T$ defined in (\ref{eq:marginal-dist}), we can obtain sample paths $x^{(i)}_{0:T}$ properly weighted with respect to the target distribution $p(x_{0:T}|\mathcal I_{0:T})$ through sequentially generating samples according to
$\mathbb Q_0,\mathbb Q_1,\cdots,\mathbb Q_T$.

Conventional SMC approaches \citep{gordon1993novel,liu1998sequential} set the forward propagation measures
$\mathbb Q_t (x_{0:t})$ to
\be \mathbb P^0_t (x_{0:t})
= p(x_{0:t}|\mathcal{I}_{0:t}),\qquad t = 0,1,\dots, T,
\label{eq:measure-Q}\ee
using only the information up to time $t$.
Under this setting, the recursion relationship of the
forward propagation measures becomes
\ben \mathbb P^0_t(x_{0:t}) \propto \mathbb P^0_{t-1}(x_{0:t-1})\cdot p(x_t\,|\, x_{0:t-1})p(\mathcal{I}_t\,|\, x_{0:t}), \een
and the incremental weight is calculated by
\ben u_t^{(i)}\propto \frac{ p(x_t^{(i)}\,|\, x_{0:t-1}^{(i)})
p(\mathcal{I}_t\,|\, x_{0:t}^{(i)})}{q(x_t^{(i)}\,|\, x_{0:t-1}^{(i)})}.\een
Here $p(x_t^{(i)}\,|\, x_{0:t-1}^{(i)})$ and $p(\mathcal{I}_t\,|\, x_{0:t}^{(i)})$
are usually specified by the model and are easy to work with.

In a constrained problem, sampling with respect to the forward sampling measure
$\mathbb P^0_t$
in (\ref{eq:measure-Q}) fails to correct the sample proactively, since it does not use any future information and constraints. It is not efficient, especially when future information
imposes strong constraints on the current state. To overcome this drawback, we propose another sequence of probability measures $\mathbb P^*_t$ which
uses part of future information to correct the Monte Carlo samples proactively.
Define $\mathbb P_t^*$ as the measure with density
\ben
 \mathbb P^*_t(x_{0:t}) = p(x_{0:t}\,|\,\mathcal I_{0:t_+}),
\qquad t = 0,1,\dots,T,
\een
where $t_+\geqslant t$ is the next time when a strong constraint is imposed after time $t$ (inclusive). If there is no strong constraint
after time $t$, we define
$t_+=t$.
Whether a constraint is "strong" depends on specific problems and is user-defined. In later sections, we will show some examples of strong constraints.


Note that $\mathbb P_t^*$
agrees with $\mathbb P_t$ and $\mathbb P^0_t$ at time $T$ since $T_+=T$ by definition.
The sequence of measures $\mathbb P_t^*$ is a compromise between the
marginal posterior measure $\mathbb P_t$ and the
forward propagation measure $\mathbb P^0_t$, where the former considers the whole information set and the latter ignores all future constraints.
When $\mathcal I_t$ is trivial,  measure $\mathbb P^*_t$ seeks the next available strong constraint $\mathcal I_{t+}$ for guidance, but not the entire future information
set as the measure $\mathbb P_t$ would require.
In most cases, the next available strong constraint $\mathcal I_{t_+}$ plays an important role in shaping the path distribution.
Hence the measure $\mathbb P^*_t$ is expected to approximate the marginal posterior measure $\mathbb P_t$ reasonably well.

To use measure $\mathbb P_t^*$, the challenges are to draw samples from
the ``optimal" proposal distribution $q(x_t\,|\, x_{0:t-1})
=\mathbb P_t^*(x_t\,|\, x_{0:t-1})=p(x_t\,|\, x_{0:t-1},\mathcal I_{0:t_+})$ and to
evaluate the incremental weights, especially when $t_+$ is far away from $t$.
Notice that $\mathbb P_t^*(x_{0:t}) \propto \mathbb P^0_t(x_{0:t})$ $p(\mathcal I_{t+1:t+}|x_{0:t})$,
where $\mathcal{I}_{t+1:t+}=\mathcal{I}_{t+1}\cap \cdots \cap \mathcal{I}_{t+}$ when $t_+\geq t+1$ and $p(\mathcal I_{t+1:t_+}\,|\, x_{0:t})=1$ when $t_+=t$.
A properly weighted sample set under measure $\mathbb P^0_t$
can be easily changed to the measure $\mathbb P_t^*$ by multiplying the weights by
$p(\mathcal I_{t+1:t+}|x_{0:t})$ or conducting a resampling step with priority scores proportional to $p(\mathcal I_{t+1:t+}|x_{0:t})$.
We choose to use the resampling approach
since it is often difficult to obtain the exact values of
$p(\mathcal I_{t+1:t+}|x_{0:t})$ except in the case
$t_+=t$ and the resampling approach is less sensitive to using the approximate
values of $p(\mathcal I_{t+1:t+}|x_{0:t})$. Specifically,
we propose to track the exact weight $w_t^{(i)}$ under
measure $\mathbb P^0_t$, but
use a resampling step with priority score \be \beta_t^{(i)}=w_t^{(i)}p(\mathcal I_{t+1:t+}\,|\, x_{0:t}^{(i)})\label{eq:priority-score}\ee
to
adjust the distribution of samples, where $p(\mathcal I_{t+1:t+}\,|\, x_{0:t}^{(i)})$ can be replaced by an approximated value.
We will discuss how to approximate $p(\mathcal I_{t+1:t+}\,|\, x_{0:t}^{(i)})$ in Section~\ref{sec:priority-score}.
Then the samples generated under $\mathbb P_t^0$ will
approximately follow measure $\mathbb P_t^*$
after resampling. We refer to this method as the constrained sequential
Monte Carlo (cSMC) method. The details of the algorithm are depicted in Figure~\ref{fig:cSMC}.

\begin{figure}[hthp]
  \begin{center}
\begin{boxedminipage}{6.5in}
\caption{Constrained Sequential Monte Carlo (cSMC) Algorithm } \label{fig:cSMC}
\it{
\begin{itemize}
\item At times $t=0,1,\cdots,T$:
\begin{itemize}

\item Propagation: For $i=1,\cdots,n$,
\begin{itemize}
\item Draw $x_t^{(i)}$ from distribution $q(x_t|x^{(i)}_{0:t-1})$ and let $x_{0:t}^{(i)}=(x_{0:t-1}^{(i)}, x_t^{(i)})$.
\item Update weights by setting
\ben
w_t^{(i)}\leftarrow w_{t-1}^{(i)}\cdot \frac{ p(x_t^{(i)}\,|\, x_{0:t-1}^{(i)})
p(\mathcal{I}_t\,|\, x_{0:t}^{(i)})}{q(x_t^{(i)}\,|\, x_{0:t-1}^{(i)})}.
\een
\end{itemize}

\item Resampling (optional):
\begin{itemize}
\item Assign a priority score $\beta_{t}^{(i)}=w_{t}^{(i)}\hat{p}(\mathcal I_{t+1:t_+}\,|\, x_{0:t}^{(i)})$
to each sample $x^{(i)}_{0:t}$,
$i=1,2,\dots, n$.
\item Draw samples $\{J_1, \dots, J_n\}$ from the set $\{1, \dots, n\}$ with replacement, with probabilities proportional to $\{\beta_{t}^{(i)}\}_{i=1,\dots, n}$.
\item Let $x_{0:t}^{*(i)}=x_{0:t}^{(J_i)}$ and $w_{t}^{*(i)}= w_{t}^{(J_i)}/\beta_{t}^{(J_i)}=1/\widehat{p}(\mathcal I_{t+1:t_+}\,|\, x_{0:t}^{(J_i)})$.
\item Return the new set $\{(x_{0:t}^{(i)}, w_{t}^{(i)})\}_{i=1,\dots, n}\leftarrow \{(x_{0:t}^{*(i)}, w_{t}^{*(i)})\}_{i=1,\dots, n}$.
\end{itemize}

\end{itemize}
\item Return the weighted sample set $\{(x_{0:T}^{(i)}, w_T^{(i)})\}_{i=1,\dots, n}$.
\end{itemize}
}
\end{boxedminipage}
\end{center}
\end{figure}

The key step in cSMC is the resampling step with priority score $\beta_t^{(i)}=w_t^{(i)}p(\mathcal I_{t+1:t+}|x_{0:t})$.
In general, resampling is done to
prevent the samples from weight collapse
\citep{kong1994sequential,liu1998sequential, liu2001monte}.
After a resampling step with resampling probabilities proportional to a set of priority scores, the samples assigned with low priority scores tend to be
replaced by those with high priority scores \citep{chen2005stopping,doucet2006efficient,fearnhead2008computational}.
In our case, we can regard the priority scores as the sampler's preferences over different sample paths. In cSMC, we use priority scores that take
future constraints into consideration. Resampling with this choice of priority scores
tends to keep paths with larger tendencies to comply with the next strong constraint, and
eliminate the unlikely paths proactively.

\begin{figure}[hthp]
\center
\includegraphics[width=0.8\textwidth]{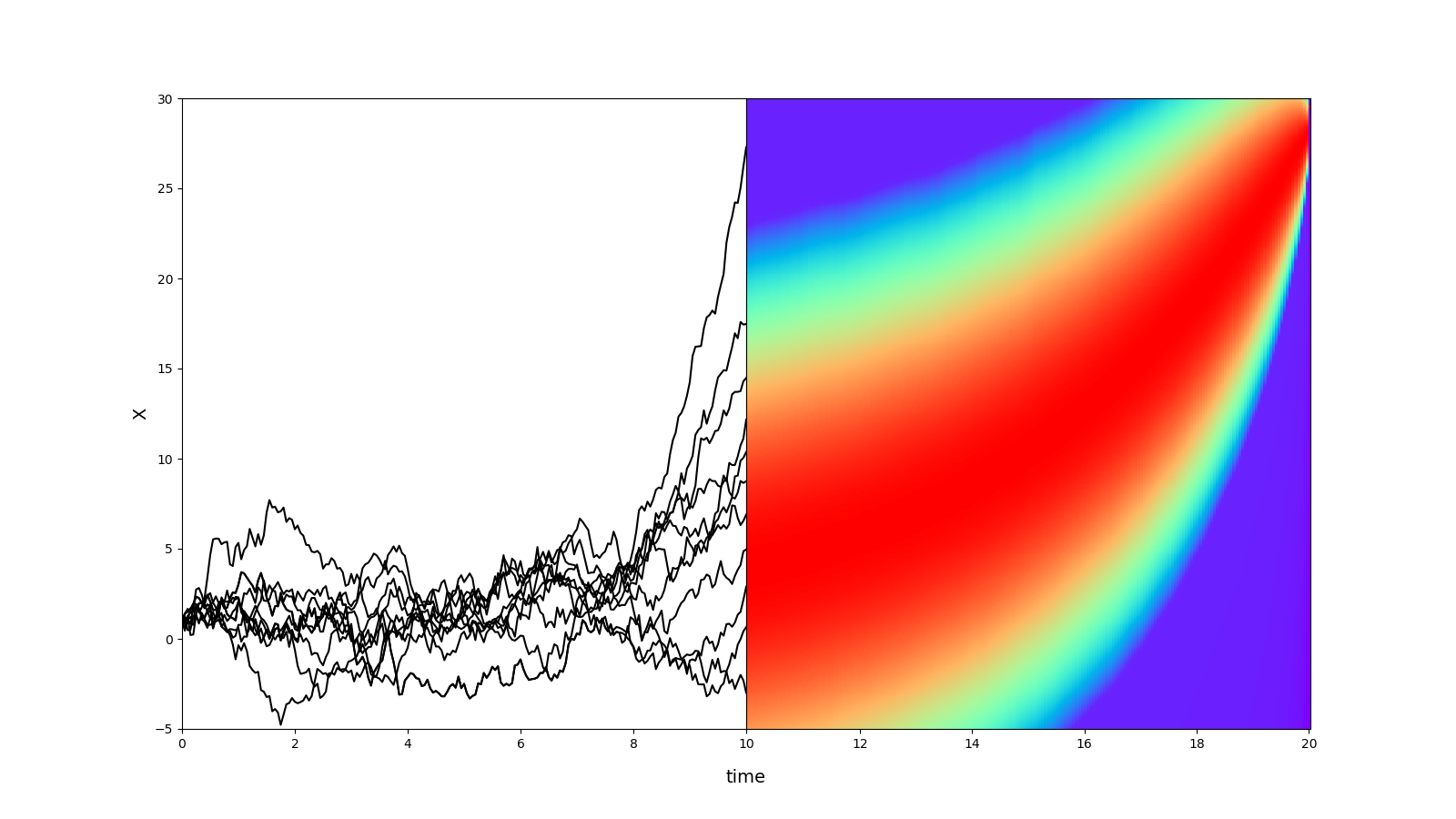}
\caption{Illustration of the resampling step at time $t=10$ in cSMC.
The left side shows several forward paths $x_{0:t}^{(i)}$ to be
resampled, and the right side shows the heatmap of $p(\cI_{t+1:t_+}\,|\, x_{0:t})$.}
\label{fig:BP}
\end{figure}

Figure \ref{fig:BP} demonstrates the resampling step at time $t=10$ in cSMC for a non-linear Markovian
stochastic process with fixed start and end points at $X_0=1$ and $X_{20}=30$.
The right side of
the figure shows the heatmap of $p(\cI_{t+1:t_+}\,|\, x_{0:t})
=p(\cI_{t+1:t_+}\,|\, x_{t})$ as a function of $t$ and $x_t$, which is
estimated by the backward pilot approach described in Section~\ref{sec:backward-pilot}. The high and low density regions of $p(\cI_{t+1:t_+}\,|\, x_{0:t})$ are colored by red and blue accordingly.
The left side of the figure shows several forward paths $x_{0:t}^{(i)}$, to be
resampled according to the priority score $\beta_t^{(i)}=w_t^{(i)} p(\cI_{t+1:t_+}\,|\, x_{0:t}^{(i)})$. The paths reaching the
low density region (blue) at $t=10$ are assigned with relatively lower priority scores and are more likely to be replaced by other paths that reaching the
high density region (red) in the resampling step.

In the proposed cSMC algorithm, we point out that
the sample set $\{(x_t^{(i)},w_t^{(i)})\}_{i=1,\cdots,n}$ obtained  at each time $t<T$ is properly weighted
with respect to $\mathbb{P}^0_t$, not the desired $\mathbb{P}_t^*$. However,
$\big\{(x_t^{(i)},w_t^{(i)}p(\mathcal I_{t+1:t+}|x_{0:t}^{(i)})\big\}_{i=1,\cdots,n}$ is properly
weighted with respect to $\mathbb{P}_t^*$. Unfortunately it is not operational since one would need to
be able to calculate $p(\mathcal I_{t+1:t+}|x_{0:t})$ precisely.


As pointed out by \cite{liu1998sequential}, conducting resampling at every time $t$ is not necessary.
It increases computational costs and introduces additional variation to the current sample set. \cite{liu1998sequential}
proposed to carry out the resampling step at a pre-determined schedule, say, performing resampling at $t=k,2k,\cdots$,
or when the effective sample size is below a certain threshold. The effective sample size (ESS) at time $t$ is defined as
\be \text{ESS}_t=\frac{n}{1+\widehat{\text{var}}(\beta_t)/ \overline{\beta}_t^2}=\frac{\left(\sum_{i=1}^n \beta_t^{(i)}\right)^2}
{\sum_{i=1}^n(\beta_t^{(i)})^2}, \label{eq:ESS}\ee
where $\overline{\beta}_t =\frac{1}{n}\sum_{i=1}^n \beta_t^{(i)}$ and
$\widehat{\text{var}}(\beta_t)=\frac{1}{n}\sum_{i=1}^n \big(\beta_t^{(i)}-\overline{\beta}_t\big)^2$.
Here the ESS$_t$ measures the variation of the priority scores $\{\beta_t^{(i)}\}_{i=1,\cdots,n}$.

\section{Special Cases} \label{sec:special-cases}

In this section, we discuss several special cases of constrained sampling problems.
All these problems can be effectively solved under the cSMC Algorithm in Figure \ref{fig:cSMC}, by specifying $\mathcal I_t$ and
the "ideal" priority score $\beta_t^{(i)}$ in (\ref{eq:priority-score}). We will discuss how to approximate $\beta_t^{(i)}$ in Section 4.


\subsection{Frequent Constrained Problems -- the State Space Model}

Consider a type of frequent constraint problems where
$x_{0:T}=(x_0,x_1,\cdots,x_T)$ is a stochastic process
governed by a forward propagation equation $p(x_{t}|x_{t-1})$ and
at each time $t$, we observe $y_t$ which is related to $x_t$ with
uncertainty. Suppose that the distribution of $y_t$ is entirely determined by $x_t$ through a conditional distribution
$p(y_t\,|\, x_t)$. Such a system is often called a state space model. The observed sequence $y_{0:T}=(y_0,\cdots,y_T)$
can be viewed as a set of frequent constraints.


This problem can be solved by sampling $x_{0:T}$ first using the forward propagation equation, and then re-weighting the paths by
$p(y_{0:T}|x_{0:T})=\prod_{t=0}^T p(y_t\,|\, x_t)$,
though it is not efficient.
The standard SISR method recursively utilizes the information
$\mathcal I_t = \{y_t\}$ during the propagation. It has been shown to be extremely useful and efficient if implemented properly. A variety of
SISR implementations are actually special cases of cSMC.

Similar to the algorithm in Figure~\ref{fig:cSMC}, the Bayesian bootstrap filter proposed in \cite{gordon1993novel}
uses $q(x_t|x_{0:t-1})=p(x_t|x_{t-1})$ for propagation and  the weights are updated by $w_t^{(i)}=w_{t-1}^{(i)}p(y_t|x_t^{(i)})$.
\cite{kong1994sequential} and \cite{liu1998sequential} adopted a proposal distribution
$q(x_t|x_{0:t-1})$ $\propto p(x_t|x_{t-1})p(y_t\,|\, x_t)$, which incorporates the current information $\cI_t=\{y_t\}$
for sampling. When the current observation contains strong information about $x_t$,  \cite{lin2005independent} proposed to
draw state samples from $q(x_t|x_{0:t-1}) \propto p(y_t\,|\, x_t)$. All above methods use $\beta_t^{(i)}=w_t^{(i)}$ for resampling,
which is the case that $t_+$ always equals $t$ in cSMC. The auxiliary particle filter (APF) proposed in \cite{pitt1999filtering} uses a different
approach. The APF
conducts resampling with the priority score $\beta_t^{(i)}=w_t^{(i)}p(y_{t+1}\,|\, x_t^{(i)})$, which is the case $t_+=t+1$ in cSMC.  They
showed that it is often more efficient to incorporate the future information $\mathcal{I}_{t+1}=\{y_{t+1}\}$  for resampling.
\cite{chen2000adaptive} and \cite{LinChenLiu13_delay}
 proposed the delayed sampling method, in which the priority score $\beta_t^{(i)}=w_t^{(i)}p(y_{t+1},\dots, y_{t+\Delta}\,|\, x_t^{(i)})$ uses future information up to a fixed delay of $\Delta$ time units. In this case, $t_+ = t + \Delta$.

\subsection{Strong Constrained Problems}\label{sec:strong-constrained-problem}

A constraint set $\mathcal{I}_{0:T}$ is said to be extremely strong if the likelihood function
$p(\mathcal{I}_{0:T}|x_{0:T})=0$ almost surely under the system dynamics $p(x_{0:T})$. That is, the constraints will never be satisfied if
we use the system dynamics to generate samples.

One example with extremely strong constraints is the diffusion bridge sampling problem
considered in \cite{pedersen1995consistency,durham2002numerical} and \cite{lin2010}.
Suppose that a continuous-time process $\{X_{\lambda}\}_{0\leq \lambda \leq \Lambda}$ is governed
by a diffusion stochastic differential equation
\be\label{eq:sde}
dX_{\lambda} = \mu(X_{\lambda},{\lambda})d{\lambda}+\sigma(X_{\lambda},{\lambda})dW_{\lambda},
\ee
where $\mu(X_\lambda,\lambda)$ and $\sigma(X_\lambda,\lambda)$ are the corresponding drift and diffusion coefficients, and
$\{W_\lambda\}_{0\leq \lambda\leq \Lambda}$ is a standard Brownian motion.
We want to generate bridge samples that connect two fixed end points $X_0=a$ and $X_\Lambda=b$.

Let $0=\tau_0<\tau_1<\cdots<\tau_{T-1}<\tau_T=\Lambda$
be a sequence of equally-spaced intermediate points and let $\delta=\tau_t-\tau_{t-1}$.
By the Euler-Maruyama method, system (\ref{eq:sde}) can be approximated by
\ben
X_{\tau_{t}}= X_{\tau_{t-1}}+\mu(X_{\tau_{t-1}}, \tau_{t-1})\delta
+\sigma(X_{\tau_{t-1}},\tau_{t-1})(W_{\tau_{t}}-W_{\tau_{t-1}})
+o_p\big(\sqrt{\delta}\big),
\een
where $o_p\big(\sqrt{\delta}\big)$ denotes the error term in discretization. High accuracy can be achieved by
increasing the number of intermediate points at the cost of additional computational burden.
In most applications, choosing the appropriate number of intermediate points is a compromise
between the discretization error and the computational efficiency.

The continuous-time stochastic process $\{X_{\lambda}\}_{0\leq \lambda\leq \Lambda}$
now is approximated by the discrete-time process $x_{0:T}$, where $x_t=X_{\tau_t}$ and
$p(x_t\,|\, x_{t-1})\sim N\big(x_{t-1}+\mu(x_{t-1},\tau_{t-1})\delta,\sigma^2(x_{t-1},\tau_{t-1})\delta\big)$.
Thus, it becomes a
sampling problem with the highly constrained target distribution $p(x_{0:T}\,|\, x_0=a, x_T=b)$.
\cite{pedersen1995consistency} used $q(x_t\mid x_{0:t-1})=p(x_t\,|\,x_{t-1})$ to generate sample paths, and
all paths are forced to connect to the fixed end point in the last step.
\cite{durham2002numerical} proposed a method based on SMC with linear interpolation as the proposal distribution.
\cite{lin2010} developed a method under the cSMC framework with $t_+=T$ for all $t$. In their method, a backward pilot approach is used
to approximate the term $p(I_{t+1:t_+}\,|\, x_t)=p(x_T=b\,|\, x_t)$ in (\ref{eq:priority-score}). \cite{lin2010} showed that using a priority score based on
the end point constraint can effectively improve the sampling efficiency.

\subsection{Systems with Intermediate Constraints }

The cSMC algorithm can also be applied to the cases with sparse intermediate constraints, which are infrequent but relatively strong.
Suppose the stochastic process $x_{0:T}$ is Markovian and is
governed by $p(x_{t+1}\,|\, x_t)$, and noisy observations come in periodically. For simplicity, we assume that
$T=KM$, and $y_k$ is a noisy measurement of $x_{kM}$ for $k=1,\cdots,K$.
Here we propose a sampling procedure for $x_{0:T}$ given the information set $y_{1:K}$ under the general cSMC framework.

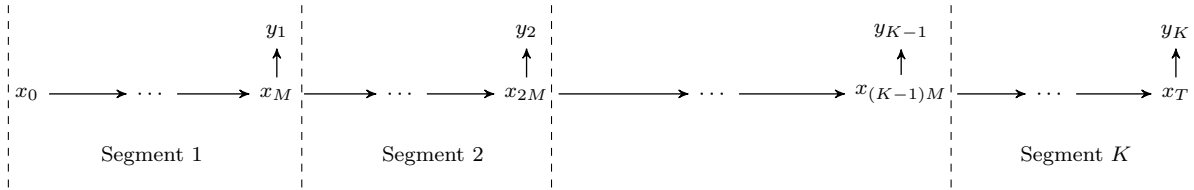
\begin{figure}[hbtp]
\center
\begin{tikzpicture}[scale = 0.83]
\node (x0) at (0,0) {\scriptsize$x_0$};
\node (x1) at (2,0) {\scriptsize$\cdots$};
\node (xM) at (4,0) {\scriptsize$x_{M}$};
\node (x2) at (6,0) {\scriptsize$\cdots$};
\node (x2M) at (8,0) {\scriptsize$x_{2M}$};
\node (x3) at (11,0) {\scriptsize$\cdots$};
\node (x3M) at (14,0) {\scriptsize$x_{(K-1)M}$};
\node (xK) at (16.4,0) {\scriptsize$\cdots$};
\node (xKM) at (18.4,0) {\scriptsize$x_{T}$};
\node (y1) at (4,1) {\scriptsize$y_{1}$};
\node (y2) at (8,1) {\scriptsize$y_{2}$};
\node (y3) at (14,1) {\scriptsize$y_{K-1}$};
\node (yK) at (18.4,1) {\scriptsize$y_K$};
\draw[->,line width = 0.5] (x0) edge (x1) (x1) edge (xM) (xM) edge (x2) (x2) edge (x2M) (x2M) edge (x3) (x3) edge (x3M) (x3M) edge (xK) (xK) edge (xKM);
\draw[->,line width = 0.5] (xM) edge (y1) (x2M) edge (y2) (x3M) edge (y3) (xKM) edge (yK);
\draw[dashed] (-0.3, -1.5) -- (-0.3,1.5) (4.4,-1.5) -- (4.4,1.5) (8.4, -1.5)--(8.4, 1.5) (14.8, -1.5)--(14.8, 1.5) (18.7, -1.5)--(18.7,1.5);
\node (s1) at (2, -1) {\scriptsize Segment 1};
\node (s2) at (6.5, -1) {\scriptsize Segment 2};
\node (sK) at (16.8, -1) {\scriptsize Segment $K$};
\end{tikzpicture}
\caption{Segmentation of a stochastic process with intermediate observations.}
\label{fig:SegChain}
\end{figure}

The intermediate observations split the whole path into $K$ segments as shown in Figure \ref{fig:SegChain}.
In the first segment, the path $(x_0,x_1,\cdots,x_M,y_1)$ can be viewed as a new system, in which $y_1$ is part of the
stochastic process and the observation equation $p(y_1\,|\, x_M)$ for $y_1$ works as the state
equation of $y_1$ conditioned on $x_M$. Under such a setting, $y_1$ is now the fixed-point
constraint at $t=M+1$. We can first draw initial samples from $q(x_0)=p(x_0)$,
then propagate to $x_M$ based on a procedure similar to sampling diffusion bridges in Section~\ref{sec:strong-constrained-problem}.
In the end, we can obtain samples from distribution $p(x_{0:M}\,|\, y_1)$. These samples of $x_{0:M}$
can be used as the initial samples for the next segment, repeated until reaching the last segment,
at which time the weighted sample set $\big\{(x_{0:T}^{(i)},w_T^{(i)})\big\}_{i=1,\dots,n}$ follows the desired distribution $p(x_{0:T}\,|\, y_1, \dots, y_K)$.

\subsection{Systems with Multilevel Constraints }\label{sec:system-multilevel}

The cSMC strategy can also be used to solve problems with multiple levels of constraints, including those with a hierarchical structure, such as one level of weak but frequent constraints and another level
of strong but infrequent constraints.
A special case is a standard state space model with two fixed endpoint constraints. Specifically, suppose a state space model is governed by the state dynamics $p(x_t\,|\, x_{t-1})$
and the observation equation $p(y_t\,|\, x_t)$. In addition, two fixed endpoint constraints are imposed on $x_{0:T}$
with $x_0=a$ and $x_T=b$. Again, we want to draw samples from the conditional distribution $p(x_{0:T}\,|\, x_0=a, y_{1:T-1},x_T=b)$.

The routine observations $y_1,\cdots,y_{T-1}$ can be viewed as a layer of weak constraints and the
fixed point constraints are viewed as a layer of strong constraints. To utilize the cSMC method, we can suppress the weak constraints layer and
define $t_+=T$ for $t>0$. That is, the priority score is chosen as
$\beta_t^{(i)} = w_t^{(i)}\,p(y_{t+1:T-1},\,x_T=b\,|\, x_t^{(i)})$ in the resampling step.

\section{Approximation of the Priority Score in cSMC}\label{sec:priority-score}

We consider the evaluation of the term $p(\mathcal I_{t+1:t_+}\,|\, x_{0:t})$ in the priority score (\ref{eq:priority-score}).
Let the time stamps of the strong constraints be $T_0<T_1<\cdots < T_K$. For the ease of presentation, we always assume that $T_0=0$
in the following. Here we omit the trivial case that $t_+=t$, in which $p(\mathcal I_{t+1:t_+}\,|\, x_{0:t})=1$, and
focus on the case that $T_{k-1}<t<T_k$ for some $k$. Then
\be p(\mathcal I_{t+1:t+}|x_{0:t})=p(\mathcal I_{t+1:T_k}|x_{0:t})=\int \cdots \int \prod_{s=t+1}^{T_K}
p(x_s\,|\, x_{0:s-1})p(\mathcal{I}_s\,|\, x_{0:s}) dx_{t+1}\cdots dx_{T_K}. \label{eq:priority-score-integration}\ee
The integrand is often well-defined by the model, but in most cases the integration does not have a closed-form solution.
In this section, we present several different
methods to approximate  $p(\mathcal I_{t+1:t_+}|x_{0:t})$.

\subsection{Optimized Parametric Priority Scores}

Based on some prior information, one may assume a parametric form for $p(\mathcal I_{t+1:t+}|x_{0:t})$.
\cite{Zhang&etal2007} and \cite{Lin&etal2008} used the cSMC approach with $t_+=T$ to generate protein conformation samples
satisfying certain residue distance constraints. The parametric functions they used to approximate $p(\mathcal I_{t+1:t+}|x_{0:t})$
are based on residue distance information of the partial chain $x_{0:t}$.

The particle efficient importance sampling (PEIS) method of \cite{Scharth&Kohn2016} uses locally optimized parametric priority scores.
Here we present PEIS under cSMC framework and notations. When the stochastic dynamic system is Markovian, that is,
\ben p(x_t\,|\, x_{0:t-1})=p(x_t\,|\, x_{t-1})\quad \text{and} \quad p(\mathcal{I}_t\,|\, x_{0:t})=p(\mathcal{I}_t\,|\, x_t),\een
for all $t$, PEIS approximates $p(\mathcal I_{t+1:T}|x_{0:t})$
and finds a series of proposal distributions $q(x_t\,|\, x_{t-1})$
close to $p(x_t\,|\, x_{t-1},\mathcal{I}_{t:T})$ for $t=0,1,\cdots,T$.
Specifically, PEIS sets
\ben q(x_t\,|\, x_{t-1}; \theta_t)=\psi_t(x_t,x_{t-1};\theta_t) / \chi_t(x_{t-1};\theta_t),\een
where $\psi_t(x_t, x_{t-1};\theta_t)$ is in a parametric family with parameter $\theta_t$, and $\chi_t(x_{t-1};\theta_t)=\int \psi_t(x_t, x_{t-1};\theta_t)\,dx_t$
is the normalizing term.
Then the importance weight at time $T$ becomes
\be \hspace{-0.5cm}w_T(x_{0:T})&=&\frac{p(x_{0:T}\,|\, \mathcal{I}_{0:T})}{q(x_0;\theta_0)\prod_{t=1}^T q(x_t\,|\, x_{t-1}; \theta_t)} \nonumber \\
&\propto& \frac{p(x_0)p(\mathcal{I}_0\,|\, x_0)\prod_{t=1}^T p(x_t\,|\, x_{t-1})p(\mathcal{I}_t\,|\, x_t)}{q(x_0;\theta_0)\prod_{t=1}^T q(x_t\,|\, x_{t-1}; \theta_t)} \nonumber \\
&\propto&\frac{p(x_0)\chi_1(x_0;\theta_1)}{\psi_0(x_0;\theta_0)}\left[\prod_{t=1}^{T-1} \frac{p(x_t\,|\, x_{t-1})p(\mathcal{I}_t\,|\, x_t)\chi_{t+1}(x_t;\theta_{t+1})}
{\psi_t(x_t, x_{t-1};\theta_t)}\right]\frac{p(x_T\,|\, x_{T-1})p(\mathcal{I}_T\,|\, x_T)}
{\psi_T(x_T,x_{T-1};\theta_T)},  \label{eq:EIS}\ee
where the initial $\psi_0(x_0;\theta_0)$ is also restricted in a parametric family.
To ensure that the difference between the target distribution $p(x_{0:T}\,|\, \mathcal{I}_{0:T})$ and the proposal distribution
$q(x_0;\theta_0)\prod_{t=1}^T q(x_t\,|\, x_{t-1}; \theta_t)$ is
small,
Equation (\ref{eq:EIS}) suggests to minimize the variation of each term in the product.
Hence, we start with an optimal $\theta_T$ that minimizes the variation of the ratio
\ben
\frac{p(x_T\,|\, x_{T-1})p(\mathcal{I}_T\,|\, x_T)}
{\psi_T(x_T,x_{T-1};\theta_T)}.
\een
Then going backward recursively,  for $t=T-1, \dots, 1$, we find $\theta_t$ that
minimizes the variation of
\ben
\frac{p(x_{t}\,|\, x_{t-1})p(\mathcal{I}_{t}\,|\, x_{t})\chi_{t+1}(x_{t};\theta_{t+1})}
{\psi_t(x_{t},x_{t-1};\theta_{t})}.
\een

The PEIS method can be easily adapted to our settings to approximate $p(\mathcal I_{t+1:t+}|x_{0:t})$.
The algorithmic steps to find the ``optimal"  parameters are presented in Figure~\ref{fig:EIS}.
We repeat the optimization procedure for each time interval from $T_{k-1}+1$ to $T_k$ to find the ``optimal"  parameter
$\theta_t$ for every $T_{k-1}<t\leq T_k$. Then
the normalizing term $\chi_{t+1}(x_t;\theta_{t+1})$ can be used as an approximation of $p(\mathcal I_{t+1:t+}|x_{0:t})$.
We note that the performance of this method greatly depends on the choice of the parametric family for $\psi_t(x_t, x_{t-1};\theta_t)$.

\begin{figure}[hthp]
  \begin{center}
\begin{boxedminipage}{6.5in}
\caption{PEIS Parameter Optimization Algorithm under cSMC Framework} \label{fig:EIS}
\it{

\begin{itemize}
\item For $k=1,\cdots,K$:

\begin{itemize}
\item Initialize the parameters $\theta_t^{[0]}$ for $t=T_{k-1}+1,\cdots,T_k$.

\item Update the parameters iteratively as follows.
\begin{itemize}
\item Generate samples $x_{T_{k-1}:T_k}^{(j)}$, $j=1,\cdots,m$, from the proposal distribution
\ben q(x_{T_{k-1}})\prod_{t=T_{k-1}+1}^{T_k} q(x_t\,|\, x_{t-1}; \theta_{t}^{[l-1]}),\een
where $q(x_{T_{k-1}})$ is a distribution close to $p(\mathcal{I}_{T_{k-1}}\,|\, x_{T_{k-1}})$.

\item Calculate the weights
\ben w_{T_k}^{(j)} =\frac{p(\mathcal{I}_{T_{k-1}}\,|\, x_{T_{k-1}}^{(j)})}{q(x_{T_{k-1}}^{(j)})}
\prod_{t=T_{k-1}+1}^{T_k} \frac{p(x_t^{(j)}\,|\, x_{t-1}^{(j)})p(\mathcal{I}_t\,|\, x_t^{(j)})}
{q(x_t^{(j)}\,|\, x_{t-1}^{(j)}; \theta_{t}^{[l-1]})} \een

\item For $t=T_k,T_k-1,\cdots, T_{k-1}+1$, solve the minimization problem
\ben &&(\theta_t^{[l]},b_t^{[l]}) = \arg\min_{\theta,\gamma} \sum_{j=1}^m w_{T_k}^{(j)}\Big\{\log \big[p(x_t^{(j)}\,|\, x_{t-1}^{(j)})p(\mathcal{I}_t\,|\, x_t^{(j)})
\chi_{t+1}(x_t^{(j)};\theta_{t+1}^{[l]})\big]\\
&&\hspace{7.5cm} - \gamma - \log \big[ \psi_t(x_t^{(j)},x_{t-1}^{(j)};\theta)\big]\Big\}^2,\een
where $\chi_{t+1}(x_t;\theta_{t+1})$ is set to a constant when $t=T_K$.
\item Stop the iteration until the parameters converge. Let  $\theta_t^*$, $t=T_{k-1}+1,T_k-1,\cdots, T_k$,
denote the converged parameters.

\end{itemize}
\end{itemize}

\item Return the estimated functions $\Big\{\widehat{p}(\mathcal{I}_{t+1:t_+}\,|\, x_{0:t})=\chi_{t+1}(x_t;\theta_{t+1}^*)\Big\}_{t=T_{k-1}+1,\cdots,T_k-1; k=1,\cdots,K}$
to compute the priority scores in Figure~\ref{fig:cSMC}.
\end{itemize}
}
\end{boxedminipage}
\end{center}
\end{figure}

\subsection{Priority Scores Based on Forward Pilots}

When it is not easy to choose an appropriate parametric family for $p(x_t,\mathcal{I}_t\,|\, x_{0:t-1})$,
we may consider to send out pilot samples to estimate the integration (\ref{eq:priority-score-integration}) by nonparametric methods. The pilot sample idea has been proposed by \cite{wang2002delayed}
and \cite{zhang2002new}, and is used for delayed estimation in SMC
\citep{LinChenLiu13_delay}.
Specifically, suppose at time $t$ we have the samples $\{(x_{0:t}^{(i)},w_t^{(i)})\}_{i=1,\cdots,n}$ properly weighted with respect to
$\mathbb{P}^0_t$. For each sample $x_{0:t}^{(i)}$, the pilot samples $\widetilde{x}_{t+1:t_+}^{(i,j)}=(\widetilde{x}_{t+1}^{(i,j)},\cdots,\widetilde{x}_{t_+}^{(i,j)})$,
$j=1,\cdots,J$, are generated from
a proposal distribution $\prod_{s=t+1}^{t_+} g(x_s\,|\, x_{0:t}^{(i)},x_{t+1:s-1})$ and are weighted by $U_t^{(i,j)}=\prod_{s=t+1}^{t_+} u_s^{(i,j)}$ with
\ben u_s^{(i,j)}=\frac{p(\widetilde{x}_s^{(i,j)}\,|\, x_{0:t}^{(i)},\widetilde{x}_{t+1:s-1}^{(i,j)})p(\mathcal{I}_{s}\,|\, x_{0:t}^{(i)},\widetilde{x}_{t+1:s}^{(i,j)})}
{g(\widetilde{x}_s^{(i,j)}\,|\, x_{0:t}^{(i)},\widetilde{x}_{t+1:s-1}^{(i,j)})}. \een
It is easy to see that $E(U_{t}^{(i,j)}\,|\, x_{0:t}^{(i)})=p(\mathcal I_{t+1:t+}|x_{0:t}^{(i)})$.
Hence we can use $\widehat{p}(\mathcal I_{t+1:t_+}\,|\, x_{0:t}^{(i)})=\frac{1}{J}\sum_{j=1}^J U_{t}^{(i,j)}$ to
approximate $p(\mathcal I_{t+1:t+}|x_{0:t}^{(i)})$.
However, the computational cost of this method is relatively high since it requires the generation of pilot samples for every path
$x_{0:t}^{(i)}$ at every time $t$.

Suppose there exists a low dimensional statistic $S(x_{0:t})$ that summarizes $x_{0:t}$ such that
\be p(x_{t}\,|\, x_{0:t-1})= p\big(x_t\,|\, S(x_{0:t-1})\big)\quad \text{and} \quad p(\mathcal{I}_t\,|\, x_{0:t})= p\big(\mathcal{I}_t \,|\, S(x_{0:t})\big)\label{eq:forward-pilot-condition} \ee
for all $t$, and suppose we have a function $\phi(\cdot)$ such that $S(x_{0:t})=\phi\big(S(x_{0:t-1}),x_t\big)$.
Then, $p(\mathcal I_{t+1:t+}\,|\, x_{0:t})$ $=p(\mathcal I_{t+1:t+}\,|\, S(x_{0:t}))$
is a function of $S(x_{0:t})$. The idea is to use a smoothing technique on the low dimensional $S(x_{0:t})$
to reduce the computational cost.
The algorithm is presented in Figure~\ref{fig:priority-score-FP}. Note that for
$U_t^{(j)}=\prod_{s=t+1}^{T_k} \widetilde{u}_s^{(j)}$ defined in Figure~\ref{fig:priority-score-FP}, we have
\ben
E(U_{t}^{(j)}\,|\, S_t^{(j)}=S)=p(\mathcal I_{t+1:t+}|S(x_{0:t})=S).
\een
Therefore, we can use $\big\{(U_{t}^{(j)}, S_t^{(j)})\big\}_{j=1,\cdots,m}$ to estimate $p\big(\mathcal I_{t+1:t+}|S(x_{0:t})\big)$
by the nonparametric histogram function
(\ref{eq:forward-pilot-smoothing}) in Figure~\ref{fig:priority-score-FP}.
We choose not to use the kernel smoothing method here in order
to control the computational cost,
because $\widehat{p}(\mathcal I_{t+1:t_+}|S(x_{0:t}))$ needs to be evaluated
for all $x_{0:t}^{(j)}, j=1,\dots, n$ and at each time $t$.
Compared with the pilot sampling method proposed in \cite{wang2002delayed}, this algorithm only need to be conducted once
to obtain $\widehat{p}\big(\mathcal{I}_{t+1:t_+}\,|\, S(x_{0:t})\big)$ for all $t$.

The accuracy of $\widehat{p}\big(\mathcal{I}_{t+1:t_+}\,|\, S(x_{0:t})\big)$ depends on the choice of the proposal
distribution $g(x_s\,|\, x_{0:t}^{(i)},$ $x_{t+1:s-1})$ to generate the pilots. Since $\mathcal{I}_{T_k}$ is a strong constraint,
when generating pilot samples from $T_{k-1}+1$ to $T_k$, we need to incorporate the information from $\mathcal{I}_{T_k}$
in the proposal distribution $g(\cdot)$, so that the pilot samples will have a reasonable large probability to satisfy the constraint $\mathcal{I}_{T_k}$.

\begin{figure}[hthp]
  \begin{center}
\begin{boxedminipage}{6.5in}
\caption{Forward Pilot Smoothing Algorithm} \label{fig:priority-score-FP}
\it{
\begin{itemize}
\item For $k=1,\cdots,K$:

\begin{itemize}
\item Initialization: For $j=1,\cdots,m$, draw samples $S_{T_{k-1}}^{(j)}$ from a proposal distribution $g(S)$
that covers the support of $S(x_{0:T_{k-1}})$.

\item For $t=T_{k-1}+1,\cdots,T_k$, draw pilot samples forwardly as follows.
\begin{itemize}
\item Generate samples $\widetilde{x}_{t}^{(j)}$ from a proposal distribution
$g(\widetilde{x}_t\,|\, S_{t-1}^{(j)})$, and calculate $S_{t}^{(j)}=\phi(S_{t-1}^{(j)},\widetilde{x}_{t}^{(j)})$
for $j=1,\cdots,m$.

\item Calculate the incremental weights
\ben \widetilde{u}_{t}^{(j)} =\frac{p(\widetilde{x}_t^{(j)}\,|\, S(\widetilde{x}_{0:t-1}^{(j)})=S_{t-1}^{(j)})
p(\mathcal{I}_t\,|\, S(\widetilde{x}_{0:t}^{(j)})=S_{t}^{(j)})}
{g(\widetilde{x}_t^{(j)}\,|\, S_{t-1}^{(j)})}, \quad j=1,\cdots,m. \een

\end{itemize}

\item For $t=T_{k-1}+1,\cdots,T_k-1$:
\begin{itemize}
\item Compute $U_{t}^{(j)}=\prod_{s=t+1}^{T_k} \widetilde{u}_s^{(j)}$ for $j=1,\cdots,m$.

\item Let $\mathcal{S}_1\cup \cdots \cup \mathcal{S}_D$ be a partition of the support of $S(x_{0:t})$.
Estimate $p(\mathcal{I}_{t+1:t_+}\,|\, x_{0:t})=p(\mathcal I_{t+1:t_+}|S(x_{0:t}))$ by
\be
f_t(S(x_{0:t}))
=\sum_{d=1}^D \xi_{t,d}
\mathbb{I}\big(S(x_{0:t})\in \mathcal{S}_d\big) \label{eq:forward-pilot-smoothing}\ee
with
\ben \xi_{t,k}=\frac{\sum_{j=1}^m U_{t}^{(j)} \mathbb{I}\big(S_{t}^{(j)}\in \mathcal{S}_d\big)}{\sum_{j=1}^m \mathbb{I}\big(S_{t}^{(j)}\in \mathcal{S}_d\big)}. \een
where $\mathbb{I}(\cdot)$ is the indicator function.

\end{itemize}

\end{itemize}

\item Return the estimated functions $\Big\{\widehat{p}(\mathcal{I}_{t+1:t_+}\,|\, x_{0:t})=f_t(S(x_{0:t}))\Big\}_{t=T_{k-1}+1,\cdots,T_k-1; k=1,\cdots,K}$
to compute the priority scores in Figure~\ref{fig:cSMC}.
\end{itemize}
}
\end{boxedminipage}
\end{center}
\end{figure}

\subsection{Priority Scores Based on Backward Pilots}\label{sec:backward-pilot}

When the stochastic dynamic system is Markovian, we can extend the backward pilot sampling method
proposed in \cite{lin2010} to the cSMC settings. In this sampling method, the pilot samples
are generated in the opposite time direction, starting from the highly constrained time point $T_k$ and propagating backward.
The algorithm is presented in Figure~\ref{fig:priority-score-BP}.

In this algorithm, the weight for the backward pilot $\widetilde{x}_{t:t_+}$ is
\ben \widetilde{w}_t = \frac{p( \widetilde{x}_{t+1:t_+},\mathcal{I}_{t+1:t_+}\,|\, \widetilde{x}_t)}{r(\widetilde{x}_{t:t_+})}, \een
where $r(\widetilde x_{t:t_+})$ is the proposal distribution to generate the backward pilots.
Taking expectation conditional on $\widetilde{x}_t$, we have
\ben E( \widetilde{w}_t \,|\, \widetilde{x}_t) &=& \int \cdots \int \frac{p( \widetilde{x}_{t+1:t_+},\mathcal{I}_{t+1:t_+}\,|\, \widetilde{x}_t)}{r(\widetilde{x}_t,\widetilde{x}_{t+1:t_+})}\,
r( \widetilde{x}_{t+1:t_+} \,|\, \widetilde{x}_t) d\widetilde{x}_{t+1:t_+}\\
&=& p(\mathcal I_{t+1:t+}|\widetilde{x}_t) / r(\widetilde{x}_t^{(j)}),\een
where $r(\widetilde x_{t+1:t_+}\,|\, \widetilde x_t)$ and  $r(\widetilde{x}_t)$  are the conditional distribution and the marginal distribution induced from $r(\widetilde x_{t:t_+})$, respectively.
Therefore,
\ben p(\mathcal I_{t+1:t+}|\widetilde{x}_t) = r(\widetilde{x}_t)E( \widetilde{w}_t \,|\, \widetilde{x}_t).\een
Again, we can use the pilot samples $\big\{(\widetilde{x}_t^{(j)},\widetilde{w}_t^{(j)})\big\}_{j=1,\cdots,m}$ to estimate $r(\widetilde{x}_t)$ and $E( \widetilde{w}_t \,|\, \widetilde{x}_t)$
by nonparametric smoothing. A histogram estimator is
\ben \widehat{p}(\mathcal I_{t+1:t+}|x_t)&=& \widehat{r}(x_t) \widehat{E}( \widetilde{w}_t \,|\, x_t)\\
&=&\sum_{d=1}^D \frac{\sum_{j=1}^m \mathbb{I}\big(\widetilde{x}_t^{(j)} \in \mathcal{X}_d\big)}{m |\mathcal{X}_d|}\mathbb{I}\big(x_t \in \mathcal{X}_d\big)
\cdot \sum_{d=1}^D  \frac{\sum_{j=1}^m \widetilde{w}_t^{(j)} \mathbb{I}\big(\widetilde{x}_t^{(j)} \in \mathcal{X}_d\big)}
{\sum_{j=1}^m \mathbb{I}\big(\widetilde{x}_t^{(j)} \in \mathcal{X}_d\big)} \,\mathbb{I}\big(x_t \in \mathcal{X}_d\big) \\
&=& \sum_{d=1}^D  \frac{\sum_{j=1}^m \widetilde{w}_t^{(j)} \mathbb{I}\big(\widetilde{x}_t^{(j)} \in \mathcal{X}_d\big)}
{m|\mathcal{X}_d|} \,\mathbb{I}\big(x_t \in \mathcal{X}_d\big),  \een
where $\mathcal X_1\cup \mathcal X_2\cup\cdots\cup\mathcal X_D$ is a partition of the support of $x_t$ and $|\mathcal{X}_d|$
denotes the volume of $\mathcal{X}_d$.

Compared with the forward pilot method, the backward pilots here are generated backward, starting from the constrained time point $T_k$.
The strong constraint $\mathcal{I}_{T_K}$ is automatically incorporated in the proposal distribution to generate $\widetilde{x}_{T_k}$
at the beginning. Hence it is often expected to have a more accurate approximation estimation of $p(\mathcal I_{t+1:t+}|x_t)$. However,
it requires the system to be Markovian to apply this method.



\begin{figure}[hthp]
  \begin{center}
\begin{boxedminipage}{6.5in}
\caption{Backward Pilot Smoothing Algorithm} \label{fig:priority-score-BP}
\it{
\begin{itemize}
\item For $k=1,\cdots,K$:

\begin{itemize}
\item Initialization: For $j=1,\cdots,m$, draw samples $\widetilde{x}_{T_k}^{(j)}$ from a proposal distribution $r(x_{T_k})$
and set $\widetilde{w}_{T_k}^{(j)}=1/ r(\widetilde{x}_{T_k}^{(j)})$.

\item For $t=T_{k}-1,\cdots,T_{k-1}+1$, draw pilot samples backward as follows.
\begin{itemize}
\item Generate samples $\widetilde{x}_{t}^{(j)}$, $j=1,\cdots,m$, from a proposal distribution
$r(\widetilde{x}_t\,|\, \widetilde{x}_{t+1}^{(j)})$.

\item Update weights by
\ben \widetilde{w}_{t}^{(j)} =\widetilde{w}_{t+1}^{(j)}\frac{p(\widetilde{x}_{t+1}^{(j)}\,|\, \widetilde{x}_{t}^{(j)})
p(\mathcal{I}_{t+1}\,|\, \widetilde{x}_{t+1}^{(j)})}
{r(\widetilde{x}_t^{(j)}\,|\, \widetilde{x}_{t+1}^{(j)})}, \quad j=1,\cdots,m. \een

\item Let $\mathcal{X}_1\cup \cdots \cup \mathcal{X}_D$ be a partition of the support of $x_t$.
Estimate $p(\mathcal{I}_{t+1:t_+}\,|\, x_{0:t})=
p(\mathcal{I}_{t+1:t_+}\,|\, x_t)$ by
\ben f_t(x_t)=\sum_{d=1}^D \eta_{t,d}
\mathbb{I}\big(x_t\in \mathcal{X}_d\big), \een
where
\ben \eta_{t,d} = \frac{1}{m |\mathcal{X}_d|} \sum_{j=1}^m \widetilde{w}_{t}^{(j)} \mathbb{I}(\widetilde{x}_t^{(j)}\in \mathcal{X}_d), \een
and $|\mathcal{X}_d|$
denotes the volume of the subset $\mathcal{X}_d$.

\end{itemize}
\end{itemize}

\item Return the estimated functions $\Big\{\widehat{p}(\mathcal{I}_{t+1:t_+}\,|\, x_{0:t})=f_t(x_t)\Big\}_{t=T_{k-1}+1,\cdots,T_k-1; k=1,\cdots,K}$
to compute the priority scores in Figure~\ref{fig:cSMC}.
\end{itemize}
}
\end{boxedminipage}
\end{center}
\end{figure}

\section{Examples}

\subsection{Computing Long-Run Marginal Expected Shortfall}

It is important to measure the systemic risk of a firm for risk control. \cite{Acharya&Engle&Richardson2012}
proposed to use the long-run marginal expected shortfall (LRMES) as a systemic risk index, which is defined as
the expected capital shortfall of a firm during a financial crisis.
Particularly, if the market index falls by 40\% in the next six months (126 trading days), it is viewed as a
financial crisis.
Let $x_{m,t}$ and $x_{f,t}$ be the daily logarithmic prices of the market and the firm at time $t$, respectively.
The LRMES of the firm is defined as
\ben \text{LRMES}= E\big(1-e^{x_{f,T}-x_{f,0}}\,|\, e^{x_{m,T}-x_{m,0}}< 0.6\big)   \een 
with $T=126$.

Following \cite{Brownlees&Engle2012} and \cite{Duan&Zhang2016}, we assume that $\{(x_{m,t},x_{f,t})\}_{t=0,1,\cdots,T}$ follows a bivariate GJR-GARCH model.
Without loss of generality, let $x_{m,0}=x_{f,0}=0$ and
\be x_{m,t}&=&x_{m,t-1}+\sigma_{m,t}\epsilon_{m,t}, \label{eq:GJR-GARCH}\\
\sigma_{m,t}^2&=& \omega_m +\big[\alpha_m+\gamma_m \mathbb{I}(\epsilon_{m,t-1}<0)\big](\sigma_{m,t-1}\epsilon_{m,t-1})^2+\beta_m \sigma_{m,t-1}^2,\nonumber \\
x_{f,t}&=&x_{f,t-1}+\sigma_{f,t}\epsilon_{f,t} = x_{f,t-1}+\sigma_{f,t}\left(\rho_{f,t}\epsilon_{m,t} + \sqrt{1-\rho_{f,t}^2}\,\xi_{f,t}\right),\nonumber \\
\sigma_{f,t}^2&=& \omega_f +\big[\alpha_f+\gamma_f \mathbb{I}(\epsilon_{f,t-1}<0)\big](\sigma_{f,t-1}\epsilon_{f,t-1})^2+\beta_f \sigma_{f,t-1}^2, \nonumber
\ee
where $\epsilon_{m,t}\sim N(0,1)$ and $\xi_{f,t}\sim N(0,1)$;
$\{\epsilon_{m,t}\}_{t=1,\cdots,T}$ and $\{\xi_{f,t}\}_{t=1,\cdots,T}$ are independent with each other and independent over time.
The time-varying correlation coefficients $\{\rho_{f,t}\}_{t=1,\cdots,T}$ are modeled
by the dynamic conditional correlation (DCC) approach. To be specified, let
$Q_{f,1},\cdots,Q_{f,T}$ be a sequence of $2\times 2$
covariance matrices satisfying
\be
Q_{f,t}&=& (1-\alpha_C-\beta_C)\overline{Q}_f + \alpha_C\left(
                                            \begin{array}{c}
                                             \sigma_{m(t-1)} \epsilon_{m(t-1)} \\
                                             \sigma_{f(t-1)}\epsilon_{f(t-1)} \\
                                            \end{array}
                                          \right)\left(
                                            \begin{array}{c}
                                             \sigma_{m(t-1)} \epsilon_{m(t-1)} \\
                                             \sigma_{f(t-1)}\epsilon_{f(t-1)} \\
                                            \end{array}
                                          \right)' + \beta_C Q_{f(t-1)}, \label{eq:DCC}
\ee
and $\rho_{f,t}$ is defined as the correlation coefficient induced by $Q_{f,t}$.

To set parameters in (\ref{eq:GJR-GARCH}) and (\ref{eq:DCC}) for our simulation, we apply the model to S\&P500 index and the stock prices of Citigroup
from January 2, 2012 to December 31, 2017. The maximum likelihood estimates of the parameters are
$\omega_m=3.35\times 10^{-6}$, $\alpha_m=3.35\times 10^{-6}$, $\gamma_m=0.152$, $\beta_m=0.858$,
,$\omega_f=4.22\times 10^{-6}$, $\alpha_f=0.0148$, $\gamma_f=0.0542$, $\beta_f=0.935$,
$\alpha_C=0.0755$, $\beta_C=0.862$, $\sigma_{m,1}=0.0113$, $\sigma_{f,1}=0.03$, $r_{f,1}=0.705$, and
\ben \overline{Q}_f=\left(
                  \begin{array}{cc}
                    \sigma_{m,1}^2 & r_{f,1}\sigma_{m,1}\sigma_{f,1} \\
                    r_{f,1}\sigma_{m,1}\sigma_{f,1} & \sigma_{f,1}^2 \\
                  \end{array}
                \right). \een

In the following,  we will use $p(\cdot)$ to denote the distribution law
under model (\ref{eq:GJR-GARCH}) and (\ref{eq:DCC}) with the parameters obtained above. If we
draw samples $\{(x_{m,0:T}^{(i)},x_{f,0:T}^{(i)},w_T^{(i)})\}_{i=1,\cdots,n}$ properly weighted with respect to
the distribution $p(x_{m,1:T},x_{f,1:T}\,|\, x_{m,0}=0,x_{f,0}=0,x_{m,T}<c)$ with $c=\log 0.6$, then
the $\text{LRMES}$ can be estimated by
\ben \frac{\sum_{i=1}^n w_T^{(i)}\left(1-e^{x_{f,T}^{(i)}}\right)}{\sum_{i=1}^n w_T^{(i)}}.\een
Notice that
\ben p(x_{m,1:T},x_{f,1:T}\,|\, x_{m,0},x_{f,0},x_{m,T}<c)
&\propto&I(x_{m,T}<c)p(x_{m,1:T},x_{f,1:T}\,|\, x_{m,0},x_{f,0})\\
&=&I(x_{m,T}<c)p(x_{m,1:T}\,|\, x_{m,0})p(x_{f,1:T}\,|\, x_{m,0:T},x_{f,0})\\
&=&I(x_{m,T}<c)\prod_{t=1}^T p(x_{m,t}\,|\, x_{m,0:t-1})\prod_{t=1}^T p(x_{f,t}\,|\, x_{m,0:t},x_{f,0:t-1}).\een
Once we obtain a set of samples $\{(x_{m,0:T}^{(i)},w_T^{(i)})\}_{i=1,\cdots,n}$  properly weighted with respect to
the distribution $p(x_{m,1:T},\,|\, x_{m,0},x_{m,T}<c)\propto I(x_{m,T}<c)\prod_{t=1}^T p(x_{m,t}\,|\, x_{m,0:t-1})$, the samples $\{x_{f,1:T}^{(i)}\}_{i=1,\dots, n}$
from $p(x_{f,1:T}\,|\, x_{m,0},x_{m,1:T}^{(i)},x_{f,0})=\prod_{t=1}^T p(x_{f,t}\,|\, x_{m,0:t},x_{f,0:t-1})$
can be easily drawn. Hence, here we only focus on sampling $x_{m,0:T}$.

The following methods are used to generate samples from the distribution $p(x_{m,1:T}\,|\, x_{m,0},x_{m,T}<c)$ and are compared.
\begin{enumerate}[label=(\arabic*)]
\item The rejection method (Rejection): We generate samples from the distribution $p(x_{m,1:T},\,|\, x_{m,0})$
without considering the constraint. The sample is accepted if $x_{m,T}<c$.
Stop sampling until we obtain $n$ sample paths satisfying $x_{m,T}^{(i)}<c$.

\item SMC with drift method (SMC):
We generate $x_{m,1:T}^{(i)}$, $i=1,\cdots,n$, based on the equation
\be x_{m,t}&=&x_{m,t-1}+\frac{c}{T}+\sigma_{m,t}\epsilon_{m,t}, \label{eq:GJR-GARCH-drift}\\
\sigma_{m,t+1}^2&=& \omega_m +\big[\alpha_m+\gamma_m \mathbb{I}(x_{m,t}-x_{m,t-1}<0)\big](x_{m,t}-x_{m,t-1})^2+\beta_m \sigma_{m,t}^2, \nonumber \ee
with $\epsilon_{m,t}\sim N(0,1)$ and $t=1,\cdots,T-1$. Here we add a drift term $\frac{c}{T}$ to the true propagation equation to force $x_{m,t}$ to have a downward trend. The samples are weighted by
\ben w_T^{(i)}=\frac{I(x_{m,T}^{(i)}<c)\prod_{t=1}^T p(x_{m,t}^{(i)}\,|\, x_{m,0:t-1}^{(i)})}{\prod_{t=1}^T q(x_{m,t}^{(i)}\,|\, x_{m,1:t-1}^{(i)})},\een
where $q(x_{m,t}\,|\, x_{m,1:t-1})$ is the conditional distribution defined by (\ref{eq:GJR-GARCH-drift}).
No resampling step will be performed in this method,
since once being resampled using the original weight under
$\mathbb P^0_t$,
the sample paths after resampling will
follow the original forward distribution $\mathbb P^0_t$
without the drift term. (Note that in the
cSMC approach, resampling is always done with a priority score that
incorporates future information, and not by the weight induced by
$\mathbb P^0_t$.)

\item The cSMC method with parametric priority score function (cSMC-PA): We consider the cSMC algorithm in Figure~\ref{fig:cSMC}.
The propagation equation (\ref{eq:GJR-GARCH}) is used as the proposal distribution $q(x_{m,t}\,|\, x_{m,1:t-1})$ for $t=1,\cdots,T-1$
to generate sample paths.
In this example, we set $t_+=T$ for all $t$. The priority score used in
the resampling step is $\beta_t^{(i)}=w_{t}^{(i)}\widehat{p}(x_{m,T}<c\,|\, x_{0:t}^{(i)})$.
Note that the value of $p(x_{m,T}<c\,|\, x_{0:t}^{(i)})$ depends on the entire historical path $x_{0:t}^{(i)}$
in this model, the algorithm in Figure~\ref{fig:EIS} is difficult to apply to obtain
the optimized parametric priority scores. Instead, the parametric function we choose here is
\ben \widehat{p}(x_{m,T}<c\,|\, x_{0:t}^{(i)}) = \Phi\big(c; x_t^{(i)}, (T-t)\overline{\sigma}_{m}^2 \big),\een
where $\Phi(c; \mu,\sigma^2)$ is the cumulative distribution function of $N(\mu,\sigma^2)$ evaluated at the value $c$ and
$\overline{\sigma}_{m}^2$ is the long-term average of $\sigma_{m,t}^2$.

\item The cSMC method with priority scores based on forward pilots (cSMC-FP): Similarly, we consider the cSMC algorithm and use
the propagation equation (\ref{eq:GJR-GARCH}) as the proposal distribution for $t=1,\cdots,T-1$ to generate sample paths.
The term $p(\mathcal I_{t+1:T}\,|\, x_{m,0:t}^{(i)})=p(x_{m,T}<c\,|\, x_{m,0:t}^{(i)})$ in the resampling priority score is estimated by forward pilots.
Although the model is not
Markovian, we have $p(x_{m,T}<c\,|\, x_{m,0:t} )=p(x_{m,T}<c\,|\, x_{m,t}, \sigma_{m,t+1})$. By treating
$(x_{m,t}, \sigma_{m, t+1})$ as a summary statistic $S(x_{m,0:t})$, the condition (\ref{eq:forward-pilot-condition})
is satisfied, and the algorithm in Figure~\ref{fig:priority-score-FP} can be applied.
Furthermore, since
$p(x_{m,T}<c\,|\, x_{m,t}, \sigma_{m,t+1})=p(x_{m,T}-x_{m,t}<c-x_{m,t}\,|\, \sigma_{m,t+1})$,
we only need to estimate the conditional cumulative distribution function
$p(x_{m,T}-x_{m,t}<\Delta \,|\,  \sigma_{m,t+1})$ for all $\Delta$. Equation (\ref{eq:GJR-GARCH-drift}) is used to generate
forward pilot samples. To save computational cost,
we use histogram estimator for $p(x_{m,T}-x_{m,t}<\Delta \,|\,  \sigma_{m,t+1})$ with partition $\mathcal{S}_{\sigma}=\cup_{d}\{0.005 (d-1)<\sigma_{m,t+1}\leq 0.005d\}$.

\end{enumerate}
In all above approaches, we force the samples to satisfy the constraint $x_{i,T}<T$ in the last step at time $T$.
To be specific, we generate $x_{m,T}^{(i)}$ from a normal distribution
$N\left(x_{m,T-1}^{(i)},\left(\sigma_T^{(i)}\right)^2\right)$  truncated by $x_{m,T}^{(i)}<c$. An efficient method to draw samples
from a truncated normal distribution can be found in the Appendix of \cite{liu1998sequential}.

\begin{figure}[!htbp]
\begin{center}
\includegraphics[width=12cm,height=9cm]{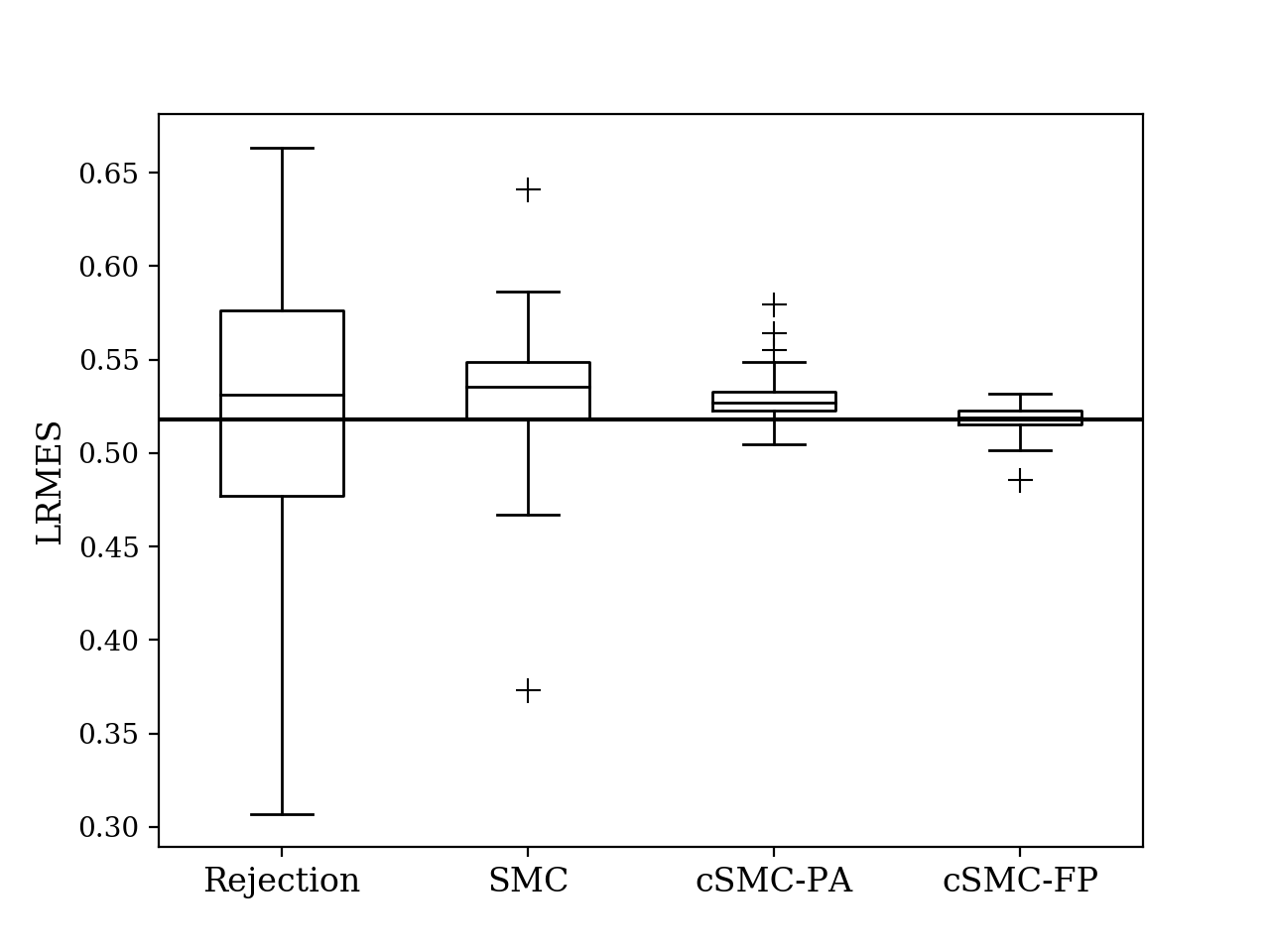}
\vspace{-0.5cm}\caption{Boxplots of 100 independent estimates of
LRMES using different methods. The horizontal line is the ``true" LRMSE estimated using
100,000 samples generated by the rejection method.}\label{Fig:LRMES-boxplot}
\end{center}
\end{figure}

The numbers of Monte Carlo samples in different methods are adjusted so that each method takes
approximately the same CPU time. More specifically, we set the surviving sample sizes to
$n=15,000$ for SMC with drift, $n=12,000$ for cSMC-PA, $n=10,000$ for cSMC-FP and $n=5$ for the rejection method. Moreover, $m=1,000$ forward pilots are sent out
to construct the resampling priority scores in cSMC-FP. In cSMC-PA and cSMC-FP, we perform resampling every 5 steps.
The acceptance rate of the rejection method is about 0.0001, due to the fact
that this is a highly constrained sampling
problem. This is the reason that we can only obtain $n=5$ surviving samples
with the same amount of computation time as the others.
Once $\{(x_{m, 0:T}^{(i)},w_T^{(i)})\}_{i=1,\cdots,n}$ is obtained, $x_{f,1:T}^{(i)}$ is sampled from $p(x_{f,1:T}\,|\, x_{m,0},x_{m,1:T}^{(i)},x_{f,0})$
for $i=1,\cdots,n$, and the corresponding LRMES is estimated.
The boxplots of 100 independent estimates of $\text{LRMES}$ using different methods are reported
in Figure~\ref{Fig:LRMES-boxplot}. It shows that cSMC-FP performs slightly better than cSMC-PA, the parametric method,
and much better than the rejection method and SMC with drift.

\begin{figure}[!htbp]
\begin{center}
\includegraphics[width=12cm,height=9cm]{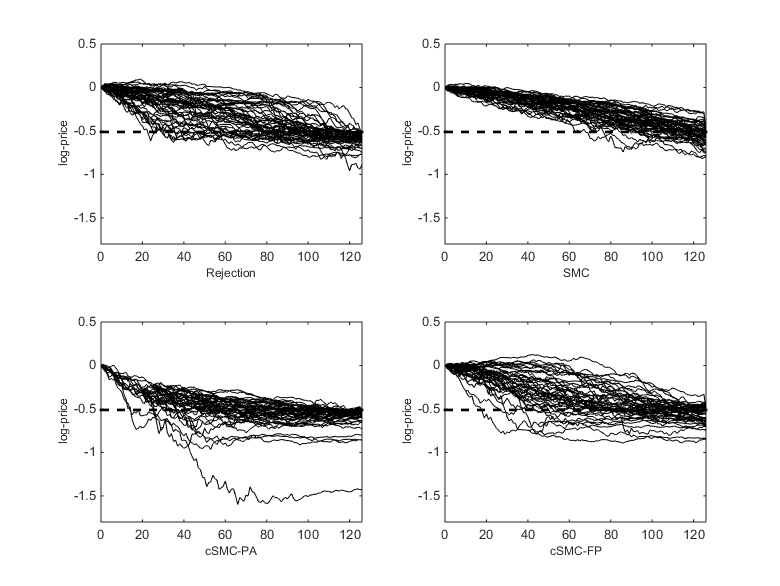}
\vspace{-0.5cm}\caption{Sample paths of
$X_{m,0:T}$ generated by different methods before weight adjustment. The horizontal line
denotes a 40\% price drop. }\label{Fig:Xm-path}
\end{center}
\end{figure}

\begin{figure}[!htbp]
\begin{center}
\includegraphics[width=12cm,height=9cm]{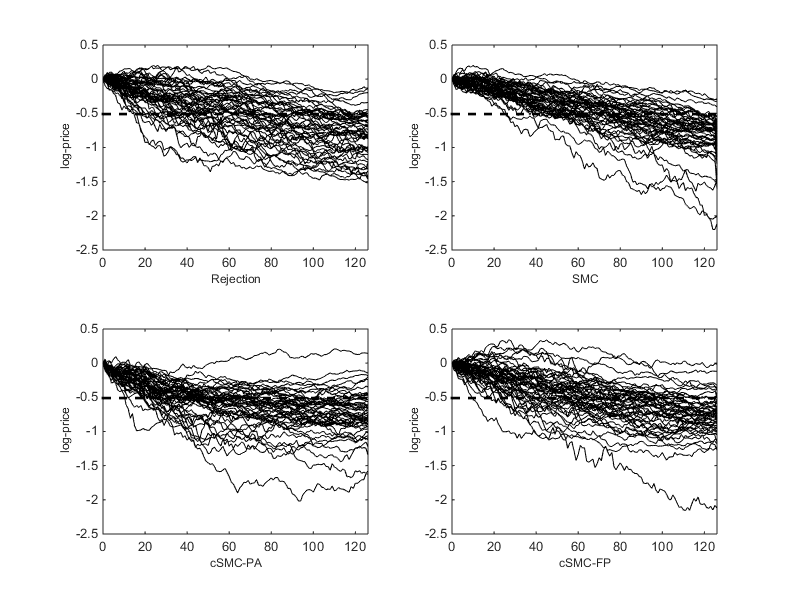}
\vspace{-0.5cm}\caption{Sample paths of
$x_{f,0:T}$ generated by different methods before weight adjustment. The horizontal line
denotes a 40\% price drop. }\label{Fig:Xf-path}
\end{center}
\end{figure}

Figure~\ref{Fig:Xm-path} and Figure~\ref{Fig:Xf-path}
plot 50 sample paths of $x_{m,0:T}$ and $x_{f,0:T}$ generated using different methods
before weight adjustment, respectively. That is, the 50 sample paths are chosen from the generated
sample set with equal probabilities, without
considering the weights.  Note that the sample paths generated by the rejection method exactly follow the true
target distribution. The figures show that, with the resampling step, cSMC-FP can
generate samples close to the true target distribution.


\subsection{System with Intermediate Observations}

Consider a diffusion process $\{X_{\lambda}\}_{0\leq \lambda\leq 90}$ governed by the following stochastic differential equation
\citep{beskos2006exact}
\ben
dX_{\lambda} = \sin (X_{\lambda}-\pi)d\lambda + dW_{\lambda},
\een
where $W_{\lambda}$ is a standard Brownian motion.
The continuous-time diffusion process
can be discretized by inserting intermediate time points with interval $\delta$. Let
$x_t=X_{t\delta}$ for $t=0,1,\cdots,T$ with $T=90/\delta$. Using the Euler-Maruyama approximation, we have
\be x_t = x_{t-1} + \delta\sin (x_{t-1}-\pi) + \varepsilon_t,\label{eq:sin_drift_dicretized}\ee
where $\varepsilon_t\sim N(0,\delta)$. We take $\delta=0.1$ in this example.

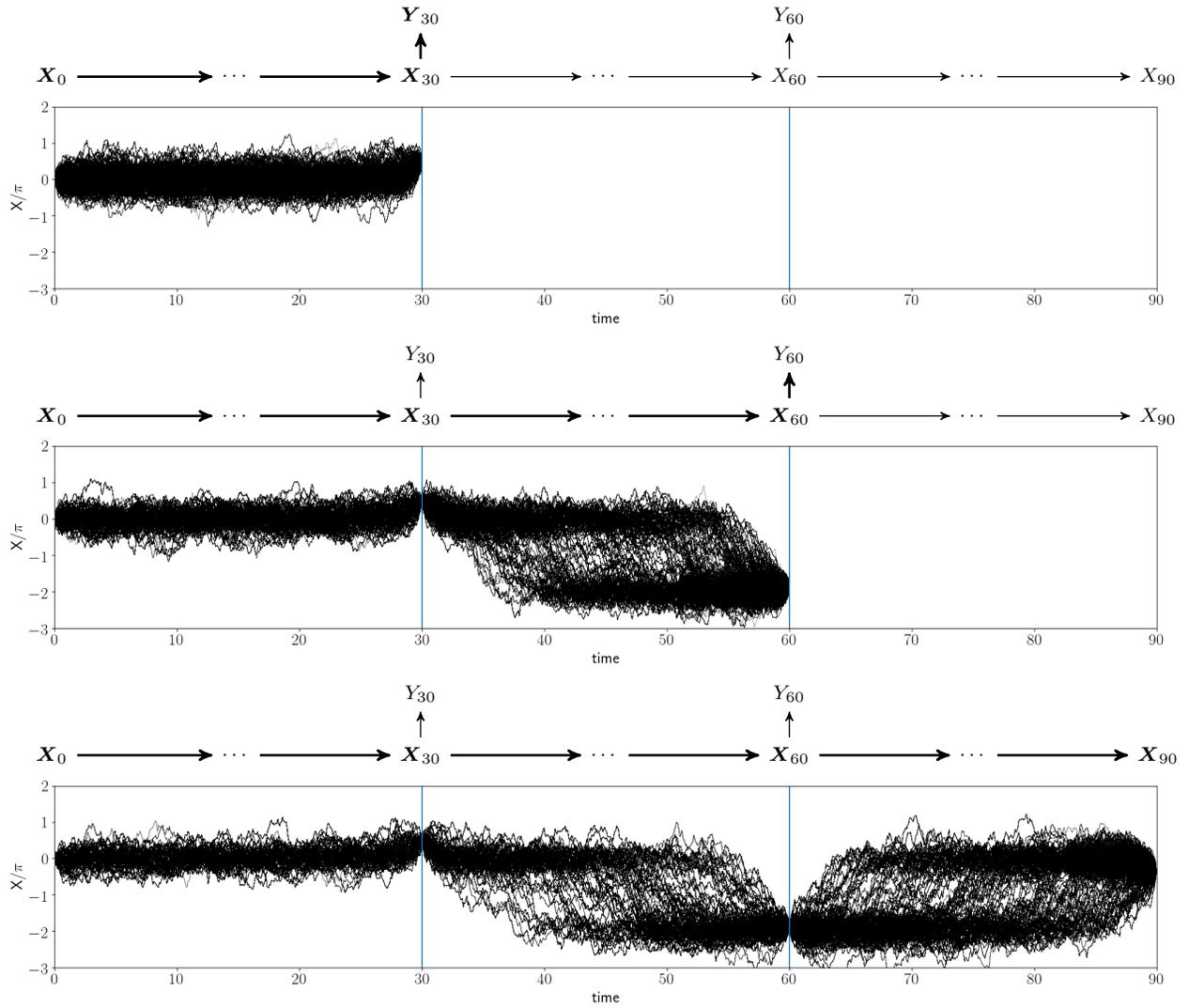
\begin{figure}[!hp]
\centering
\hspace*{8pt}
\begin{tikzpicture}[scale = 0.87]
\node (x0) at (0,0) {\scriptsize$\bX_0$};
\node (x15) at (3,0) {\scriptsize$\cdots$};
\node (x30) at (6,0) {\scriptsize$\bX_{30}$};
\node (x45) at (9,0) {\scriptsize$\cdots$};
\node (x60) at (12,0) {\scriptsize$X_{60}$};
\node (x75) at (15,0) {\scriptsize$\cdots$};
\node (x90) at (18,0) {\scriptsize$X_{90}$};
\node (y30) at (6,1) {\scriptsize$\bY_{30}$};
\node (y60) at (12,1) {\scriptsize$Y_{60}$};
\draw[->,line width = 0.5]  (x30) edge (x45) (x45) edge (x60) (x60) edge (x75) (x75) edge (x90);
\draw[->,line width = 0.5]  (x60) edge (y60);
\draw[->,line width = 1.0] (x0) edge (x15) (x15) edge (x30) (x30) edge (y30);
\end{tikzpicture}
\includegraphics[width=\textwidth]{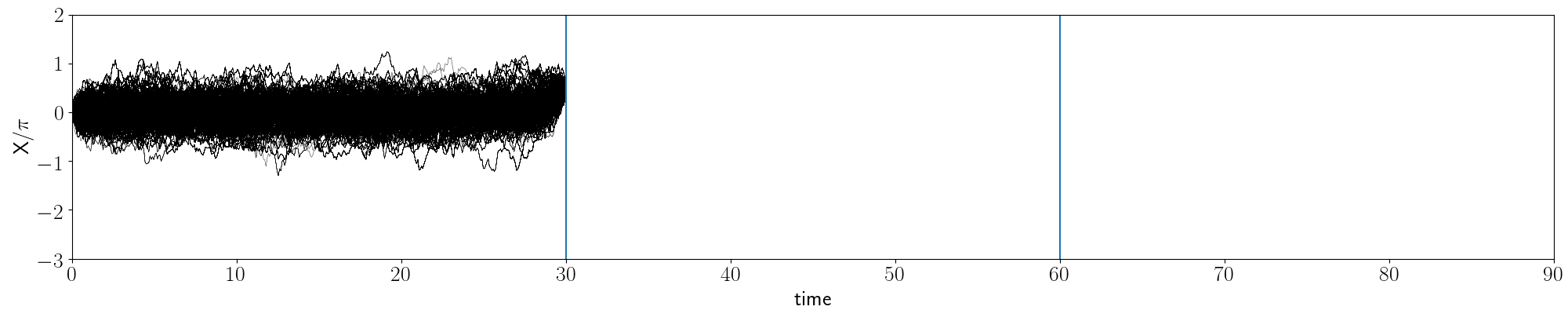}

\hspace*{8pt}
\begin{tikzpicture}[scale = 0.87]
\node (x0) at (0,0) {\scriptsize$\bX_0$};
\node (x15) at (3,0) {\scriptsize$\cdots$};
\node (x30) at (6,0) {\scriptsize$\bX_{30}$};
\node (x45) at (9,0) {\scriptsize$\cdots$};
\node (x60) at (12,0) {\scriptsize$\bX_{60}$};
\node (x75) at (15,0) {\scriptsize$\cdots$};
\node (x90) at (18,0) {\scriptsize$X_{90}$};
\node (y30) at (6,1) {\scriptsize$Y_{30}$};
\node (y60) at (12,1) {\scriptsize$Y_{60}$};
\draw[->,line width = 0.5](x60) edge (x75) (x75) edge (x90);
\draw[->,line width = 0.5] (x30) edge (y30);
\draw[->,line width = 1.0] (x60) edge (y60)  (x0) edge (x15) (x15) edge (x30) (x30) edge (x45) (x45) edge (x60) ;
\end{tikzpicture}
\includegraphics[width=\textwidth]{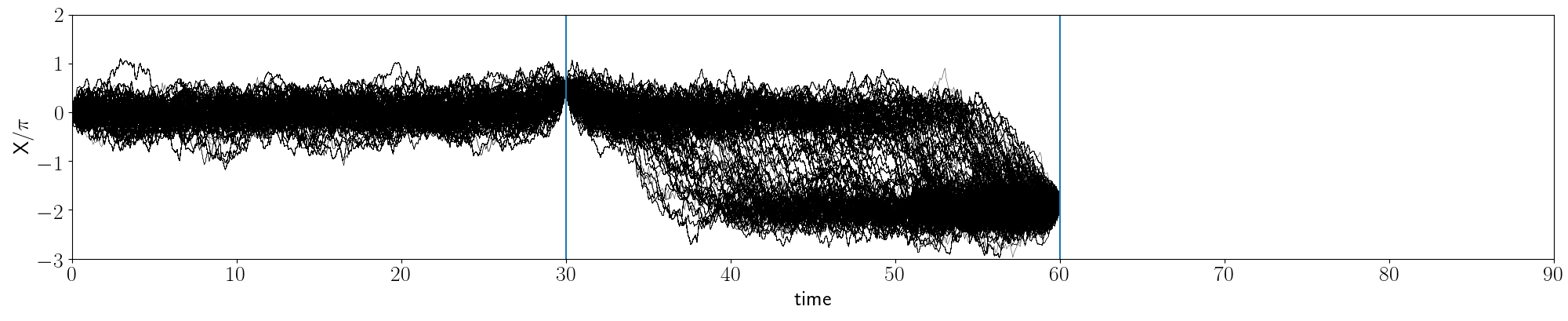}

\hspace*{8pt}
\begin{tikzpicture}[scale = 0.87]
\node (x0) at (0,0) {\scriptsize$\bX_0$};
\node (x15) at (3,0) {\scriptsize$\cdots$};
\node (x30) at (6,0) {\scriptsize$\bX_{30}$};
\node (x45) at (9,0) {\scriptsize$\cdots$};
\node (x60) at (12,0) {\scriptsize$\bX_{60}$};
\node (x75) at (15,0) {\scriptsize$\cdots$};
\node (x90) at (18,0) {\scriptsize$\bX_{90}$};
\node (y30) at (6,1) {\scriptsize$Y_{30}$};
\node (y60) at (12,1) {\scriptsize$Y_{60}$};
\draw[->,line width = 0.5] (x30) edge (y30) (x60) edge (y60);
\draw[->,line width = 1.0]  (x0) edge (x15) (x15) edge (x30) (x30) edge (x45) (x45) edge (x60) (x60) edge (x75) (x75) edge (x90) ;
\end{tikzpicture}
\includegraphics[width=\textwidth]{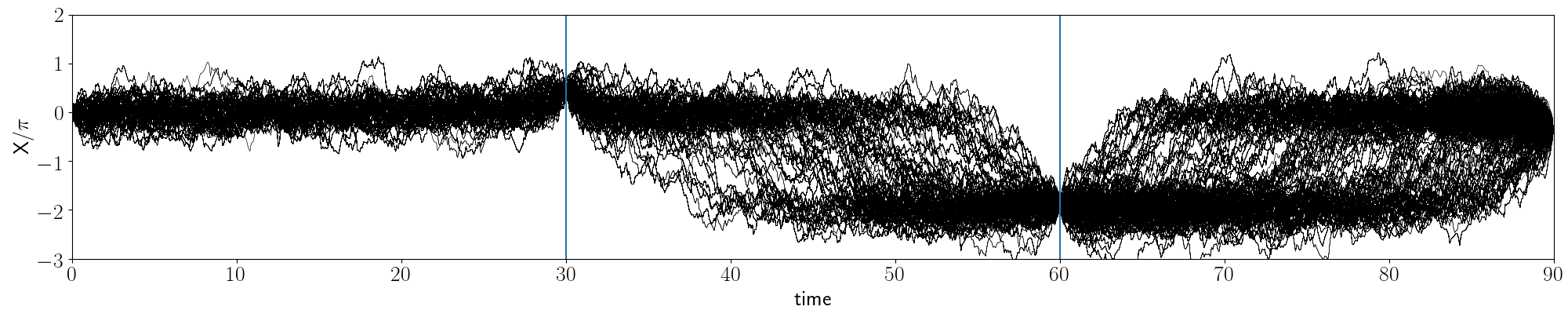}
\caption{Illustration of the segmental sampling procedure.}
\label{fig:Segmental}
\end{figure}

In this simulation study, two noisy observations of $X_{\lambda}$ are made at times $\lambda=30$ and $\lambda=60$. That is,
\ben
Y_{30}\sim N(X_{30},\sigma^2)\quad \text{and}\quad Y_{60}\sim N(X_{60},\sigma^2).
\een
We also fix the two endpoints at $X_0=a$ and $X_{90}=b$. The discretized time points $T_0=0$, $T_1=30/\delta$, $T_2=60/\delta$ and $T_3=90/\delta$ are
considered to have strong constraints.
The cSMC-BP method is applied to generate sample paths of $x_{0:T}$ conditional on
the constraints $(X_0,Y_{30},Y_{60},X_{90})$. That is, we use cSMC in Figure~\ref{fig:cSMC} to generate samples, and the backward pilot smoothing algorithm in
Figure~\ref{fig:priority-score-BP} is used to compute the resampling priority scores.
We take equation
(\ref{eq:sin_drift_dicretized}) as the proposal distribution
in generating forward paths.
The backward pilots are generated according to the algorithm in Figure \ref{fig:priority-score-BP}
with the proposal distribution $r(\widetilde x_t \mid \widetilde x_{t+1})\sim
N\big(\widetilde x_{t+1} - \delta \sin( \widetilde x_{t+1}-\pi), \delta\big)$.
Resampling is conducted dynamically when
the ESS in (\ref{eq:ESS}) is less than $0.3n$. In this example, the time line is split into three segments.
The segmental sampling procedure is demonstrated in Figure \ref{fig:Segmental}.

In the first experiment, we set
$X_0 = 0$, $Y_{30}=1.49$, $Y_{60}=-5.91$ and $X_{90}=-1.17$.
Note that this process
shows a jump behavior among the stable levels at $X_{\lambda}=2k\pi$, $k=0,\pm 1, \pm 2,\cdots$ \citep{lin2010}.
The four observations correspond to the stable levels 0, 0, $-2\pi$ and 0 accordingly.
The process is likely to fluctuate around the stable level 0 during the first period. Then, it jumps
to stable level $-2\pi$ in the second period and eventually jumps back to stable level 0 in the third period.

\begin{figure}[hthp]
\centering
\includegraphics[width = 0.8\textwidth]{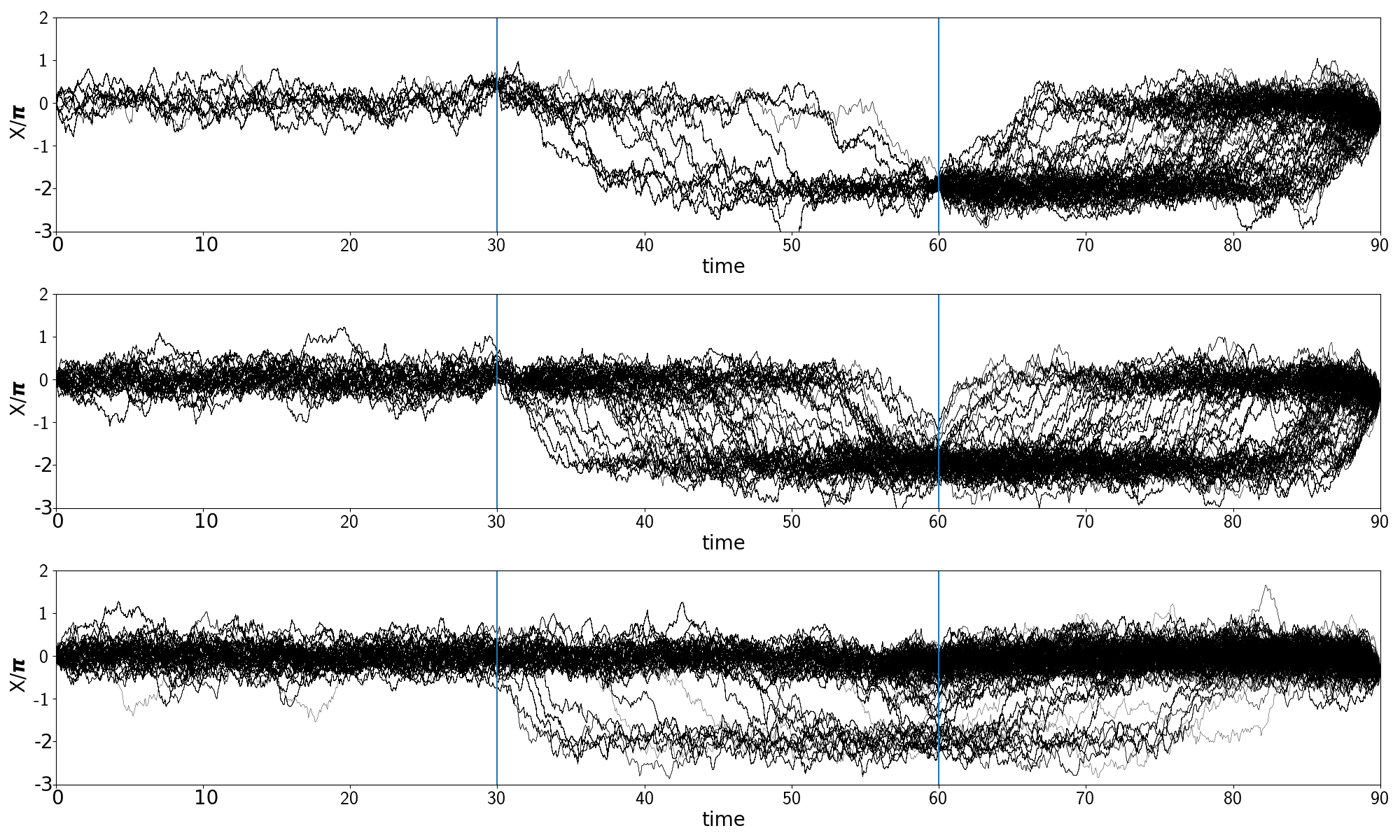}
\caption{Sampled paths before weight adjustment for $\sigma = 0.01$ (top panel), $\sigma=1.0$ (middle panel) and $\sigma=2.0$ (bottom panel)  when $X_0 = 0$, $Y_{30}=1.49$, $Y_{60}=-5.91$ and $X_{90}=-1.17$}
\label{fig:path1}
\end{figure}

\begin{figure}[hthp]
\centering
\includegraphics[width = 0.8\textwidth]{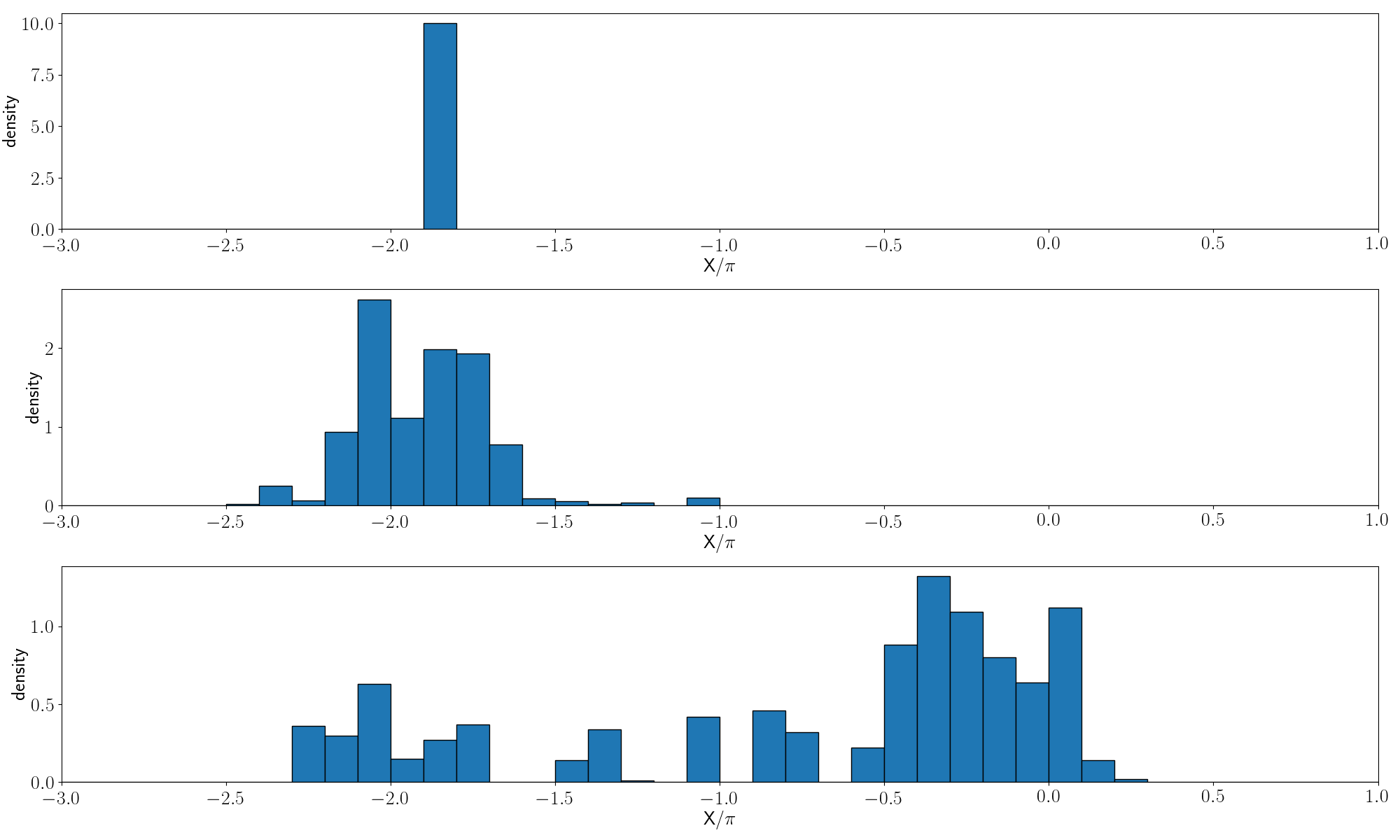}
\caption{Histogram of the marginal samples of $X_{60}$ before weight adjustment for $\sigma = 0.01$ (top panel), $\sigma=1.0$ (middle panel) and $\sigma=2.0$ (bottom panel) when $X_0 = 0$, $Y_{30}=1.49$, $Y_{60}=-5.91$ and $X_{90}=-1.17$.}
\label{fig:marginal1}
\end{figure}

Three levels of measurement errors for the observations $Y_{30}$ and $Y_{60}$ are investigated:
$\sigma = 0.01$ for very accurate observations, $\sigma = 1$ for moderate accurate observations and $\sigma = 2$ for untrusted observations.
Note that in this experiment we fix the observations $Y_{30}$ and $Y_{60}$
but changes the underlying assumption of their distributions to reflect
the strength and accuracy of the observations.
A total number of $1,000$ forward paths are generated, and $300$  backward pilots are used to estimate the resampling priority scores.
Figure \ref{fig:path1} plots the generated sample paths before weight adjustment for each level of error.
Figure \ref{fig:marginal1} shows the histogram of the marginal samples of $X_{60}=x_{60/\delta}$ before weight adjustment,
which is obtained from the generated sample set $\{x_{0:T}^{(i)}\}_{i=1,\cdots,n}$
without considering the weights.
It can be seen that when the observations are accurate ($\sigma = 0.01$),
the two observations act like fixed-point constraints that force all sample paths to pass through the observations.
When the observation error is large ($\sigma = 2$), a high proportion of sample paths remains at the original stable level
while only a small proportion of paths is drawn towards the observations.
The moderate error case ($\sigma = 1$) is a compromise between these two cases.
The marginal distributions of $X_{60}$ show clear differences in the above three cases.
Samples from all three levels of error retain the jumping nature of underlying process and the
cSMC-BP approach is capable of dealing with different levels of observational errors.


\begin{figure}[hthp]
\centering
\includegraphics[width = 0.8\textwidth]{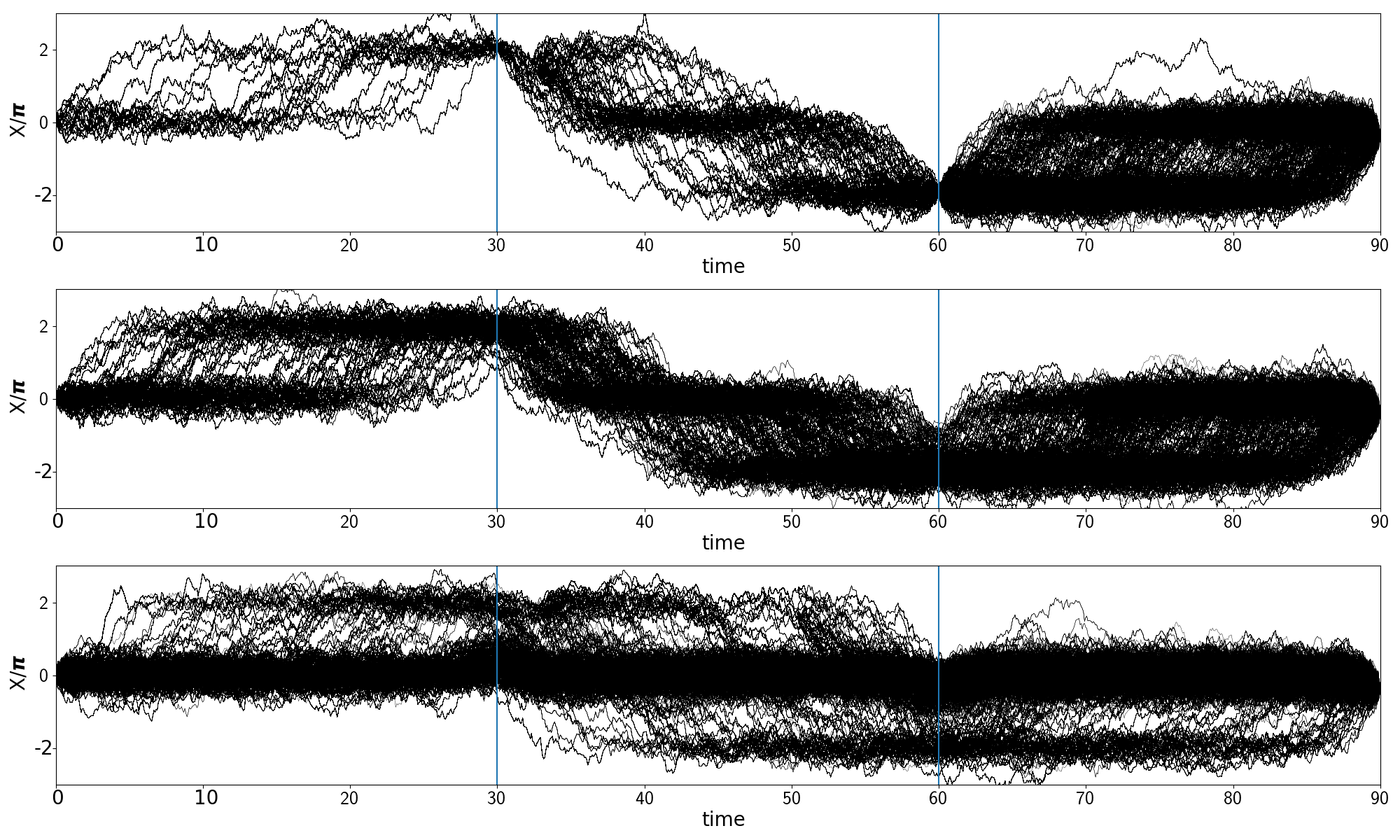}
\caption{Sampled paths before weight adjustment for $\sigma = 0.01$ (top panel), $\sigma=1.0$ (middle panel) and $\sigma=2.0$ (bottom panel)
when $X_0 = 0$, $Y_{30}=6.49$, $Y_{60}=-5.91$ and $X_{90}=-1.17$.}
\label{fig:path2}
\end{figure}

\begin{figure}[hthp]
\centering
\includegraphics[width = 0.8\textwidth]{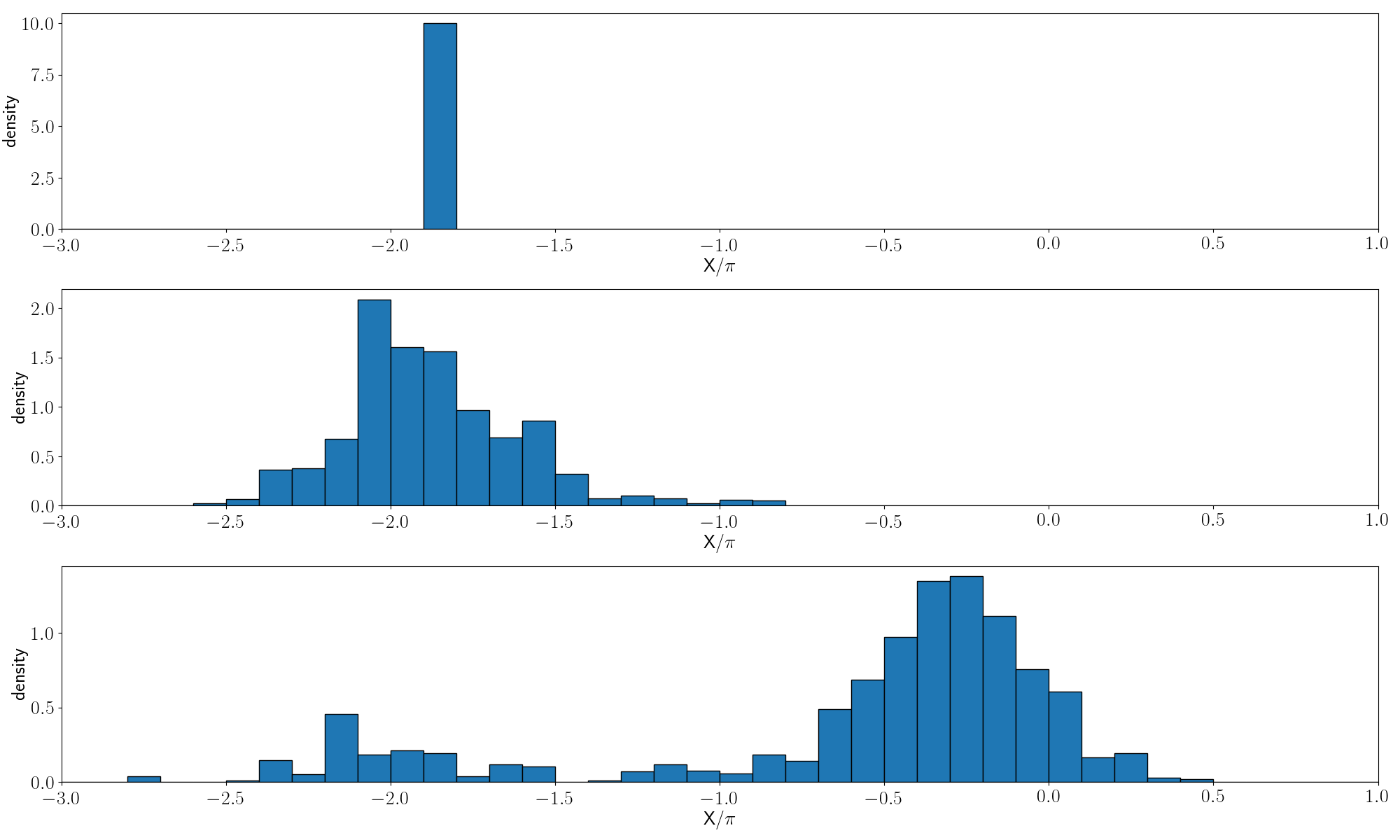}
\caption{Histogram of the marginal samples of $X_{60}$ before weight adjustment for
$\sigma = 0.01$ (top panel), $\sigma=1.0$ (middle panel) and $\sigma=2.0$ (bottom panel) when $X_0 = 0$, $Y_{30}=6.49$, $Y_{60}=-5.91$ and $X_{90}=-1.17$.}
\label{fig:marginal2}
\end{figure}

Next, we use the same settings as above except that setting $Y_{30}=6.49$. Now
the four observations $X_0=0$, $Y_{30}=6.49$, $Y_{60}=-5.91$ and $X_{90}=-1.17$ correspond to the stable levels 0, $2\pi$, $-2\pi$ and $0$, respectively.
Since $Y_{30}$ and $Y_{60}$ differ by a gap of two stable levels,
this is a very rare event. In this case, the Monte Carlo sample size is increased to 5,000 in order to overcome the degeneracy and to capture the rare event.
Sample paths before weight adjustment and histograms of $X_{60}$ samples for different levels of error are shown in Figures~\ref{fig:path2} and Figures~\ref{fig:marginal2},
respectively.
In the large error case ($\sigma = 2$), most samples are concentrated around the stable level 0.
As the error level decreases, the observation induced constraints become stronger, hence more sample paths are drawn towards the observations.
Those figures provide the evidence that the priority scores estimated by the backward pilots are effective for different error levels under this extreme setting.


\subsection{Sampling Constrained Trading Paths}

In asset portfolio management, the optimal trading path problem is
a class of
optimization problems which typically maximizes certain utility function
of the trading path \citep{markowitz1959portfolio}. This optimization problem is often complicated, especially when
trading costs are considered. \cite{kolm2015multiperiod} turned such an optimization problem
into a state space model and explored Monte Carlo methods to numerically solve it.

More specifically, let $x_{0:T}=(x_0,x_1,\dots, x_T)$ be a trading path where $x_t$ represents the holding position of an asset
in shares at time $t$.
In practice, a starting position $x_0$ and a target end
position $x_T$ are often imposed for optimal execution of a large order with
minimum market impact. Without loss of generality, we impose two endpoints at $x_0=0$
and $x_T=0$, respectively.  Then it becomes
an optimization problem to maximize the utility function
\be \label{eq:optimization}
u(x_{0:T})=-\sum_{t=1}^T c_t(x_t-x_{t-1})-\sum_{t=1}^{T-1} h_t(y_t-x_t)
\ee
given $x_0=0$ and $x_T=0$,
where $(y_1,y_1,\dots, y_{T-1})$ is a predetermined optimal trading path in an ideal world without trading costs, typically obtained by
maximizing the risk-adjusted expected return under
the Markowitz mean-variance theory \citep{markowitz1959portfolio}.
Here $c_t(\cdot)$ is the trading cost function and $h_t(\cdot)$ stands for the utility loss due to
the departure of the realized path from the ideal path. An emulating state space model can be implemented with
the state equation $p(x_t\,|\, x_{t-1})\propto \exp\{- c_t(x_t-x_{t-1})\}$ and the observation equation
$p(y_t\,|\, x_t)\propto \exp\{- h_t(y_t-x_t)\}$. The joint posterior distribution of
such a state space model is
\be p(x_{1:T-1}\,|\, x_0,y_{1:T-1},x_T)&\propto & \prod_{t=1}^T p(x_t\,|\, x_{t-1})\prod_{t=1}^{T-1} p(y_t\,|\, x_t) \nonumber \\
&\propto& \exp\left\{-\left[\sum_{t=1}^T c_t(x_t-x_{t-1})+\sum_{t=1}^{T-1} h_t(y_t-x_t)\right]\right\}.  \label{eq:optimization2} \ee
Thus, it is a state space model with fixed point constraints as described in Section~\ref{sec:system-multilevel}.

Following \cite{kolm2015multiperiod}, we set $T=20$. The ideal trading path is given by
\ben y_t = 25\exp\{-(t+1)/8\}-40\exp\{-(t+1)/4\}. \een
The trading cost function and the utility loss due to tracking error are set to
\be
c_t(x_t-x_{t-1}) = \frac{1}{2\sigma_x^2}\big[ (x_t-x_{t-1})^2+2\alpha|x_t-x_{t-1}|\big]\quad\text{and}\quad
h_t(y_t-x_t) = \frac{1}{2\sigma_y^2}(y_t-x_t)^2, \label{eq:utility-specification}
\ee
respectively, where $\sigma_x^2=0.25$ and $\sigma_y^2=1$.
Here the trading cost is assumed to be a quadratic function of the trade size $|x_t-x_{t-1}|$, and
$\alpha$ is a non-negative constant related to volatility and liquidity of the asset \citep{kyle2011market}, which we will specify in the following.

It can be seen that maximizing the utility function (\ref{eq:optimization}) is equivalent to find the
maximize-a-posterior (MAP) path of distribution (\ref{eq:optimization2}). We use a two-step method to find the
optimal trading path. First, we draw samples from the highly constrained conditional distribution (\ref{eq:optimization2}) with the settings
specified in (\ref{eq:utility-specification}). Then we discretize the space of $x_t$, $t=1,\cdots,T-1$, based on the generated sample paths.
The Viterbi algorithm  \citep{viterbi1967error, forney1973viterbi}
is applied to find an optimal path that maximizes the utility function (\ref{eq:optimization}) within the discretized state spaces.
In general, the closer the generated sample paths are to the optimal one, the better trading path
the Viterbi algorithm will produce.

We investigate two cases of $\alpha$ in (\ref{eq:utility-specification}): $\alpha=0$ and $\alpha=0.5$.
In both cases, we compare the performance of cSMC-BP with a standard SMC. The state equation $p(x_t\,|\, x_{t-1})\propto \exp\{- c_t(x_t-x_{t-1})\}$
is used to generate forward paths in both methods. However, the standard SMC uses $\beta_t^{(i)}=w_t^{(i)}$ as the resampling priority scores,
but cSMC-BP uses $\beta_t^{(i)}=w_t^{(i)}\widehat{p}(y_{t+1:T-1},x_T\,|\, x_t^{(i)})$ estimated by the backward pilot method
in Figure \ref{fig:priority-score-BP} for resampling, which takes the future information into account.
The backward pilots are generated from the proposal distribution
$$r(\widetilde{x}_t|\widetilde{x}_{t+1})\propto \exp\left\{-\frac{1}{2\sigma_x^2}\big[ (\widetilde x_t-\widetilde x_{t+1})^2+2\alpha| \widetilde x_t-\widetilde x_{t+1}|\big]\right\}.$$
We use $m=300$ backward pilots and generate $n=2,000$ forward sample paths from cSMC-BP. For the purpose of comparison, the standard SMC draws
$n=2,300$ forward paths such that both methods have a similar computational cost.
In both methods, a resampling step is conducted when the ESS in (\ref{eq:ESS}) is less than $0.3n$.

\subsubsection{Case 1: $\alpha=0$}

\begin{figure}[hthp]
\centering
\includegraphics[width=0.4\textwidth]{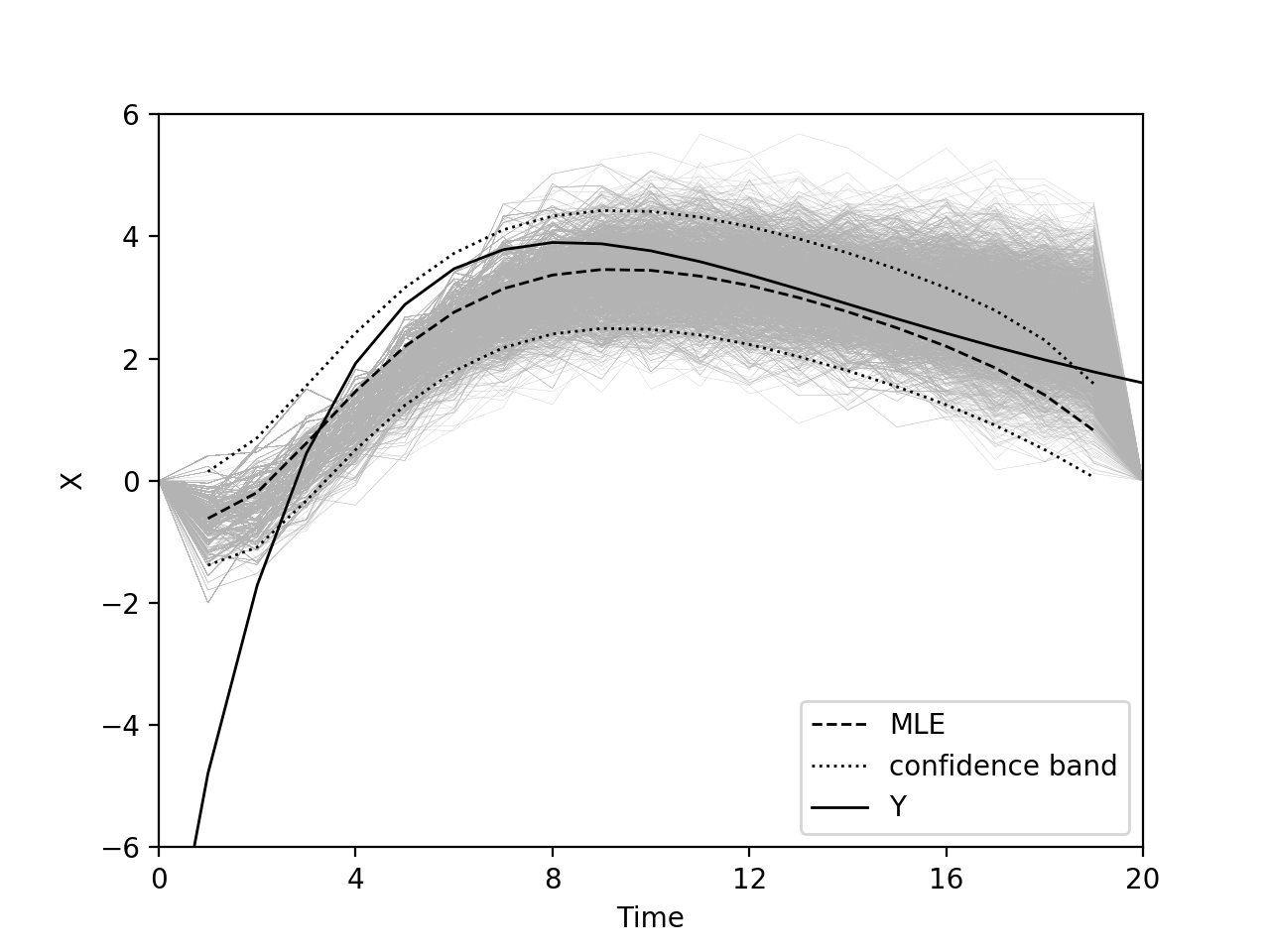}
\includegraphics[width=0.4\textwidth]{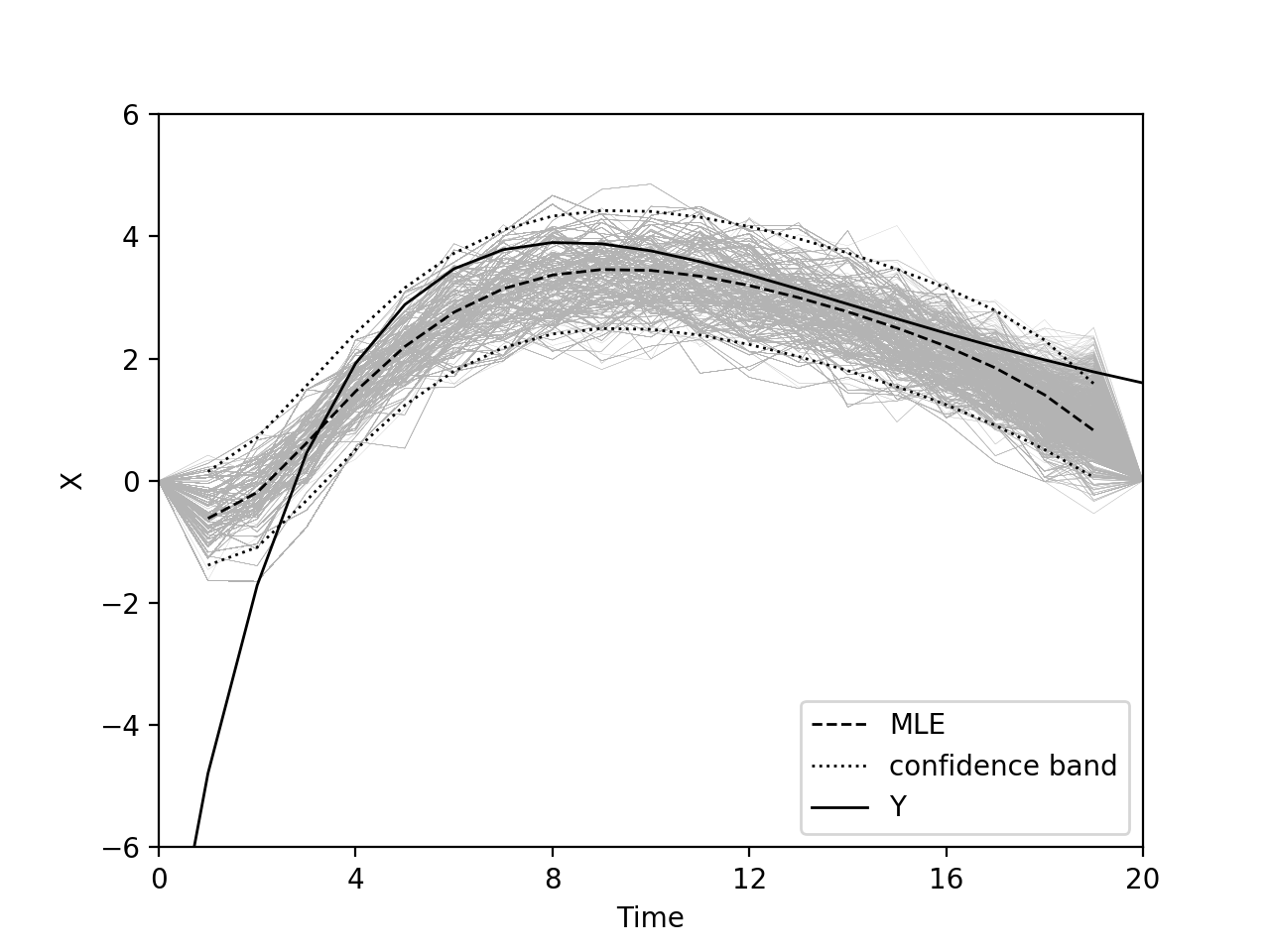}
\caption{Sample paths from the standard SMC method (left panel)
and from the cSMC-BP method (right panel) before weight adjustment when $\alpha=0$.
}
\label{fig:trade1}
\end{figure}

It can be seen that the state space model is linear and Gaussian when $\alpha=0$. Hence, the Kalman filter \citep{kalman1960new} can be applied to obtain
an exact optimal solution.
The sample paths generated from the standard SMC and cSMC-BP before weight adjustment, along with the exact optimal path and the
95\%  point-wise confidence intervals
obtained by the Kalman filter are plotted in Figure \ref{fig:trade1}.
The samples from the standard SMC in the left panel have a much larger variance and most of them lie outside the 95\% confidence region, while most samples from cSMC-BP in the right panel
stay within the 95\% confidence region. In cSMC-BP,
the backward pilots bring the information from the future and guide the forward sample paths by resampling. On the other hand, without using any future information,
the standard SMC sampler propagates blindly and suffers a large
divergency between the sampling distribution and the
target distribution in the end.

\begin{figure}[hthp]
\centering
\includegraphics[width=0.45\textwidth]{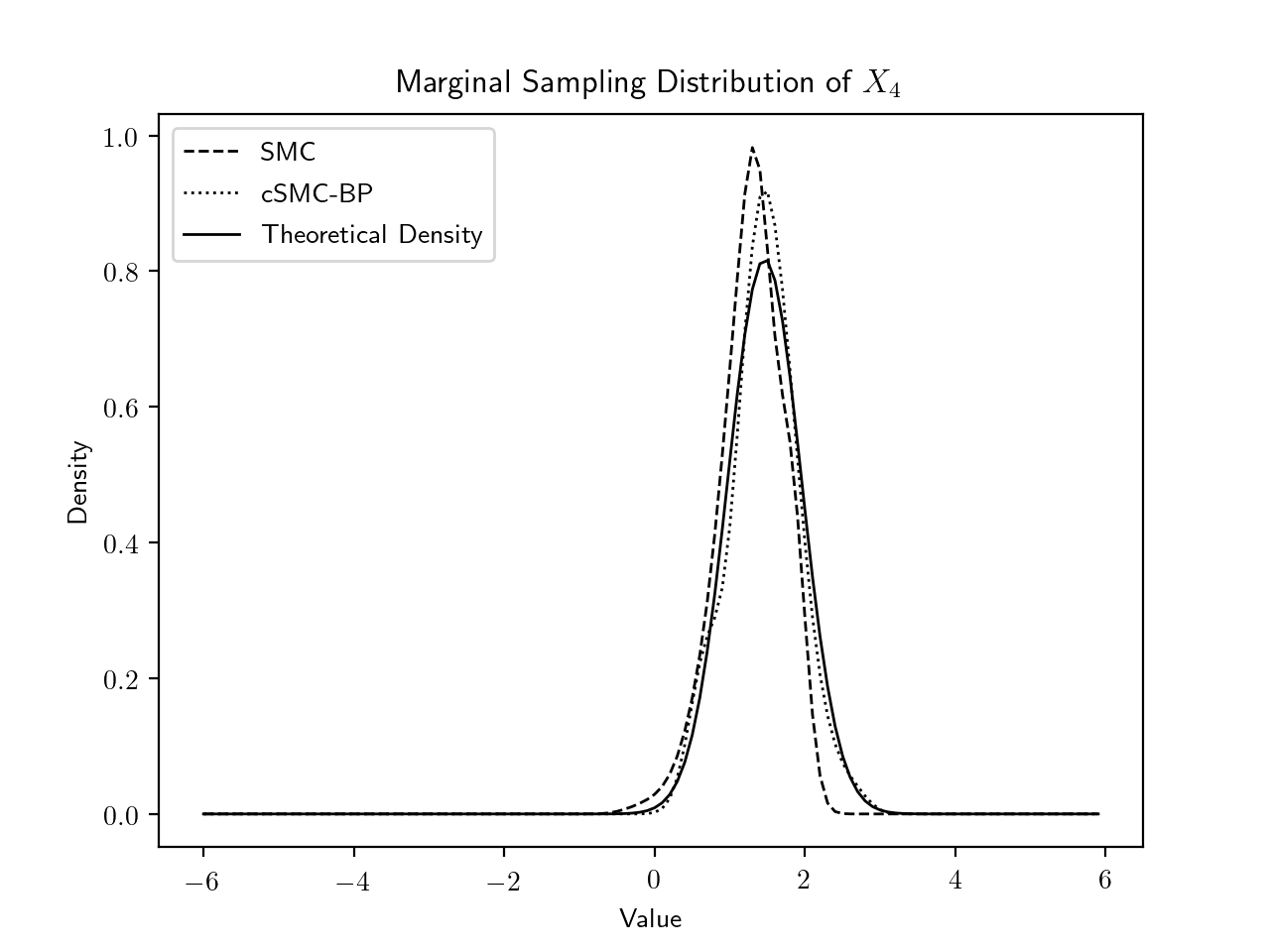}
\includegraphics[width=0.45\textwidth]{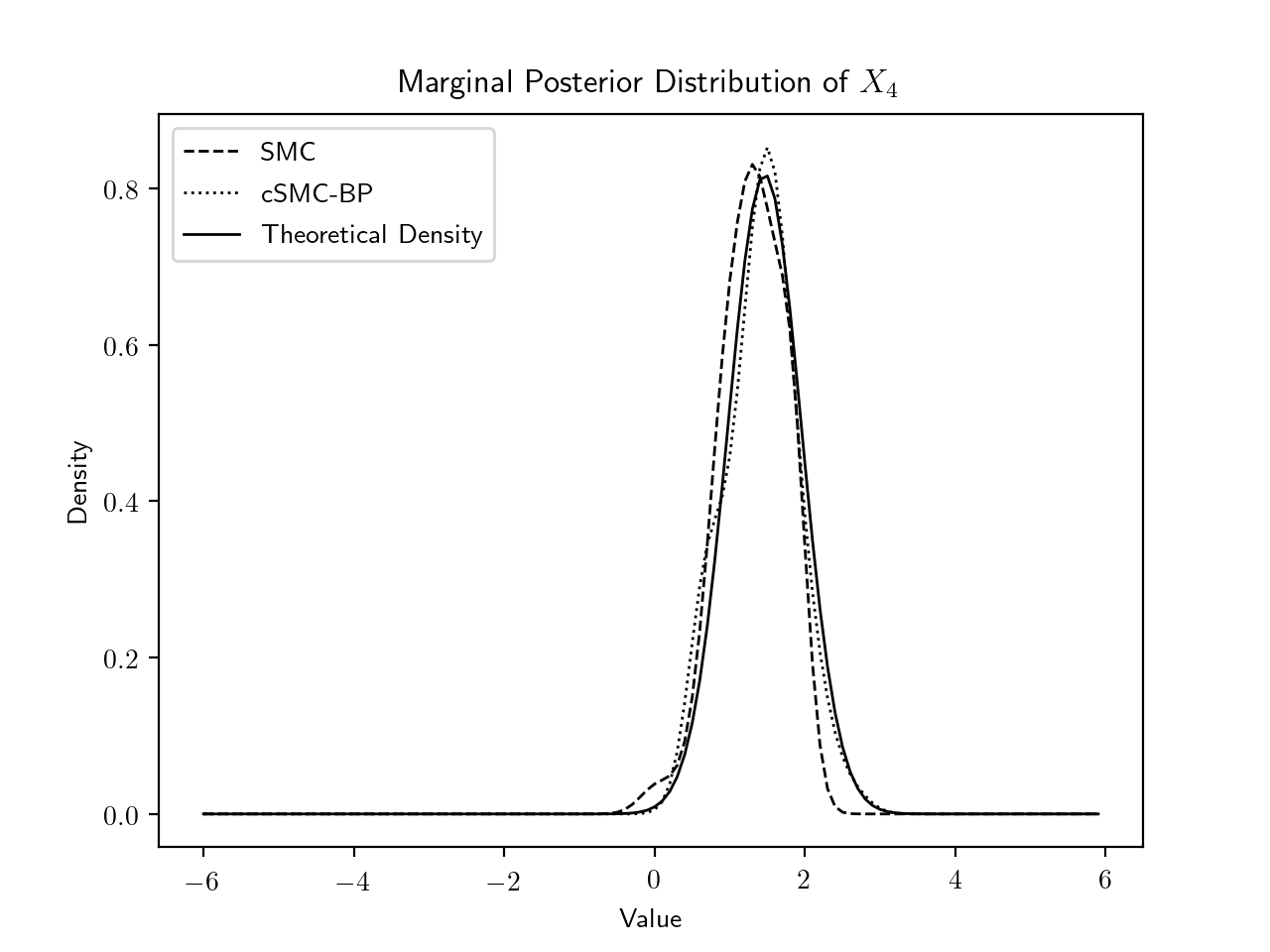}\\
\includegraphics[width=0.45\textwidth]{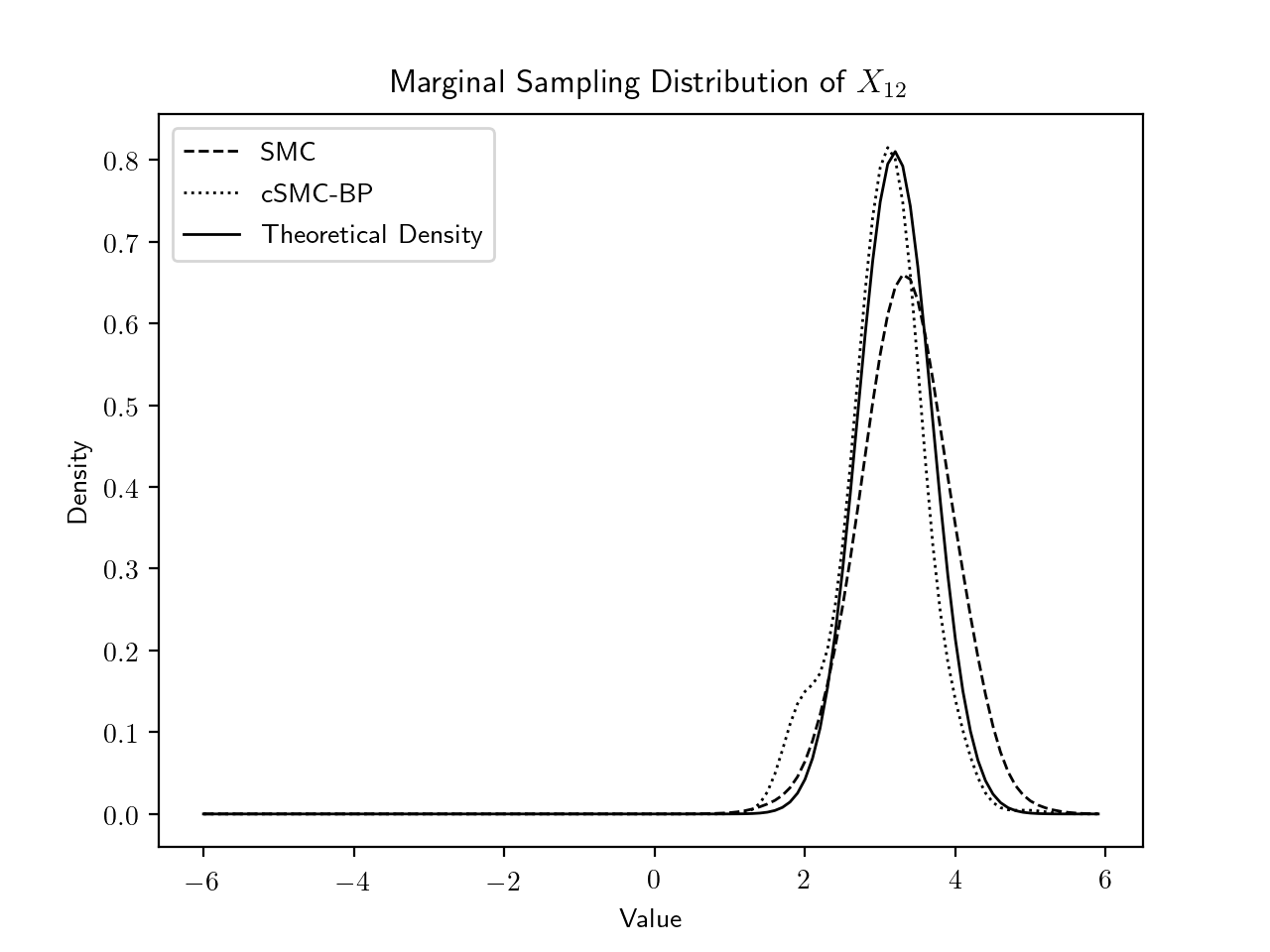}
\includegraphics[width=0.45\textwidth]{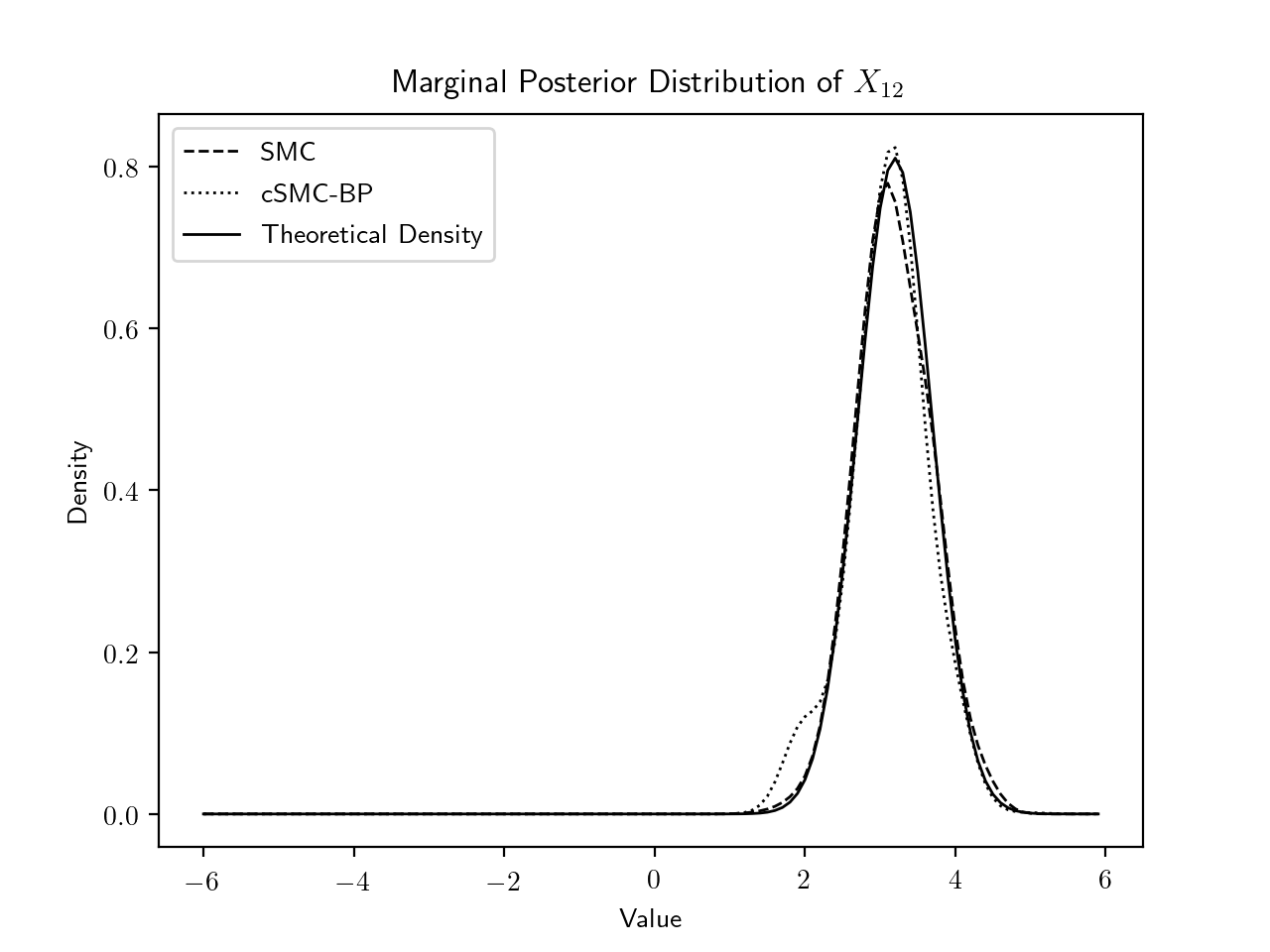}\\
\includegraphics[width=0.45\textwidth]{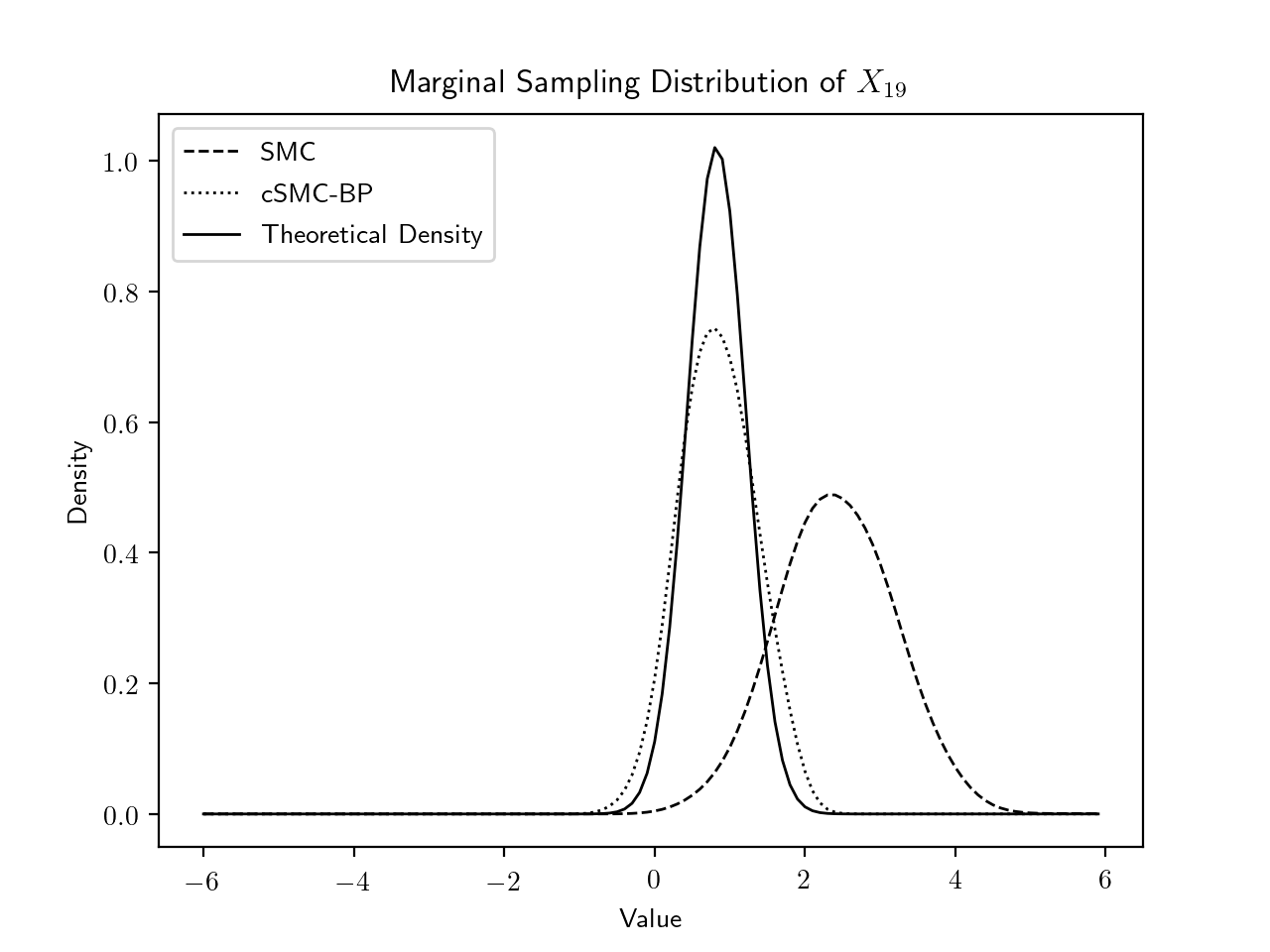}
\includegraphics[width=0.45\textwidth]{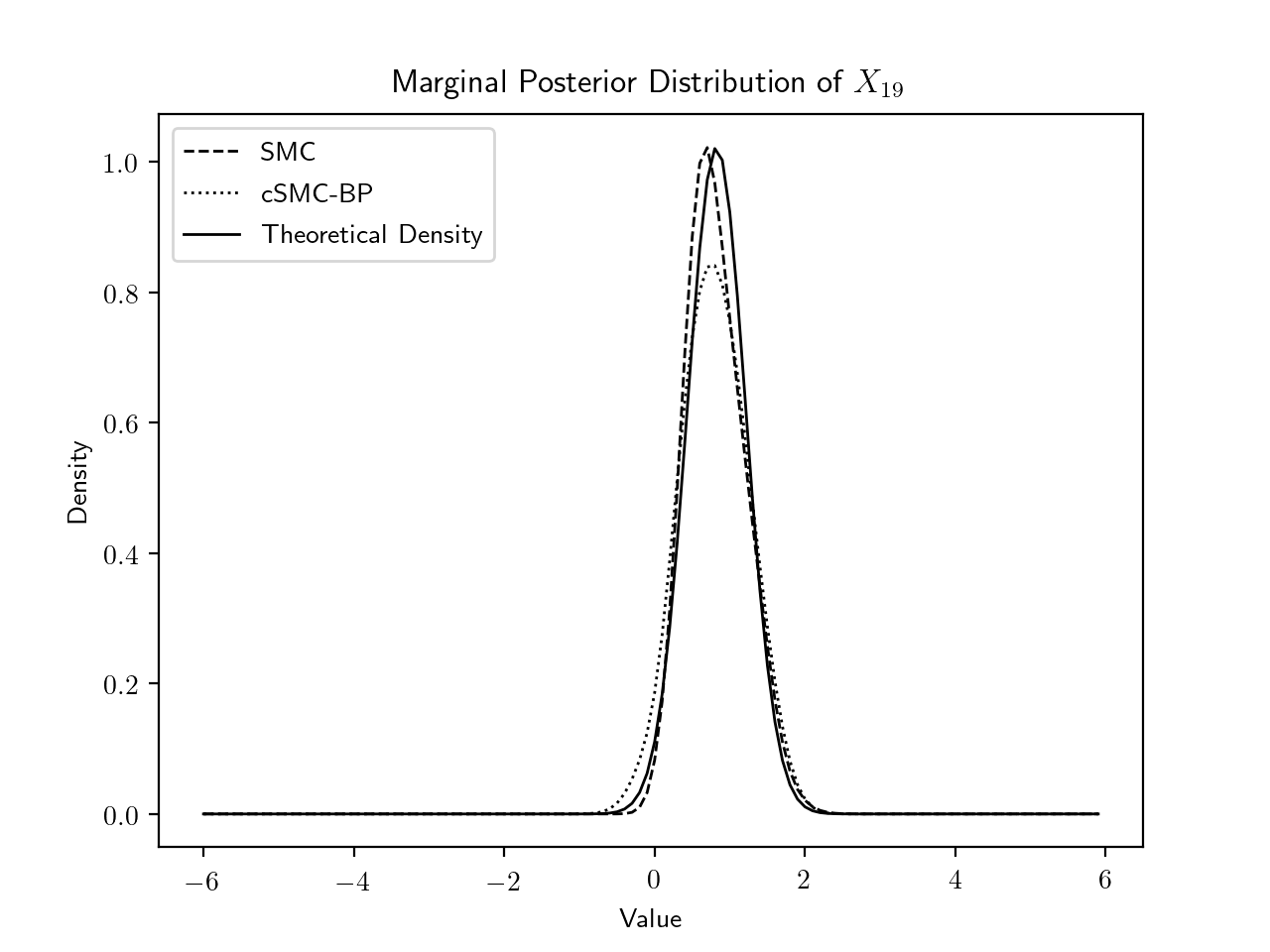}
\vspace{-0.5cm}\caption{ Marginal densities of the samples generated by
SMC and cSMC-BP before weight adjustment (left column) and after weight adjustment (right column) at
time $t=4$ (row 1), $t=12$ (row 2)  and $t=19$ (row 3) when $\alpha=0$. }
\label{fig:Marginal}
\end{figure}

Figure \ref{fig:Marginal}
shows the marginal densities of the samples generated by
the standard SMC and cSMC-BP before weight adjustment (left column) and after weight adjustment (right column) at time $t=4, 12, 19$. Both methods
produce properly weighted samples,
as the marginal densities for the samples after weight adjustment are close to the
true one. However, the sampling distribution
for $x_{19}$ before weight adjustment under the standard SMC method has a large divergence from the true distribution,
which results in a low efficiency for inference.


\begin{figure}[hthp]
\center
\includegraphics[width = 0.4\textwidth]{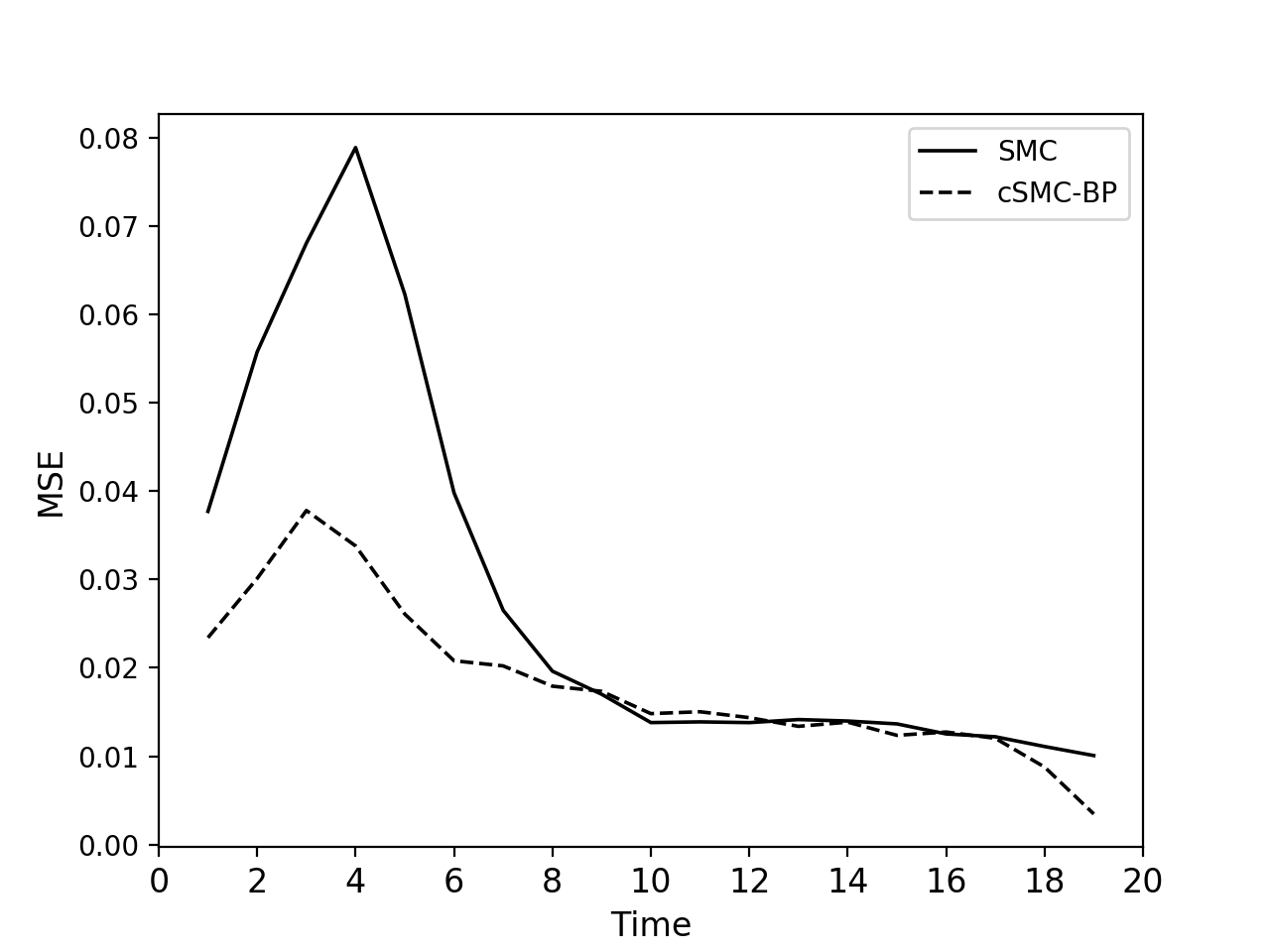}
\caption{ Mean squared error curves for SMC and cSMC-BP when $\alpha=0$.}
\label{fig:mse1}
\end{figure}

Figure \ref{fig:mse1} reports the mean squared errors (MSE) defined by
\be \text{MSE}(t)= \frac{1}{L}\sum_{l=1}^L \left[ \widehat{E}^{[l]}(x_t\,|\, x_0,y_{1:T-1},x_T) - E(x_t\,|\, x_0,y_{1:T-1},x_T) \right]^2,\label{eq:MSE}\ee
where $E(x_t\,|\, x_0,y_{1:T-1},x_T)$ is the true conditional mean obtained from the Kalman filter, and $\widehat{E}^{[l]}(x_t\,|\, x_0,y_{1:T-1},x_T)$
is the conditional mean estimated by SMC or cSMC-BP in the $l$-th replication. We use $L=1,000$ replications to compute the MSE's.
It shows that in the period $8\leqslant t\leqslant 17$ where the fixed points have limited impacts, SMC and cSMC-BP have similar performance.
But in the period $1\leqslant t\leqslant 7$ where the observation $y_t$ changes over time dramatically, cSMC-BP results in a smaller MSE than SMC
as the future information is incorporated in its resampling step.
In the period $t=18$ and $19$
where the end point constraint takes effect, the cSMC-BP approach also has a smaller MSE.

\subsubsection{Case 2: $\alpha=0.5$}

\begin{figure}[hthp]
\centering
\includegraphics[width=0.4\textwidth]{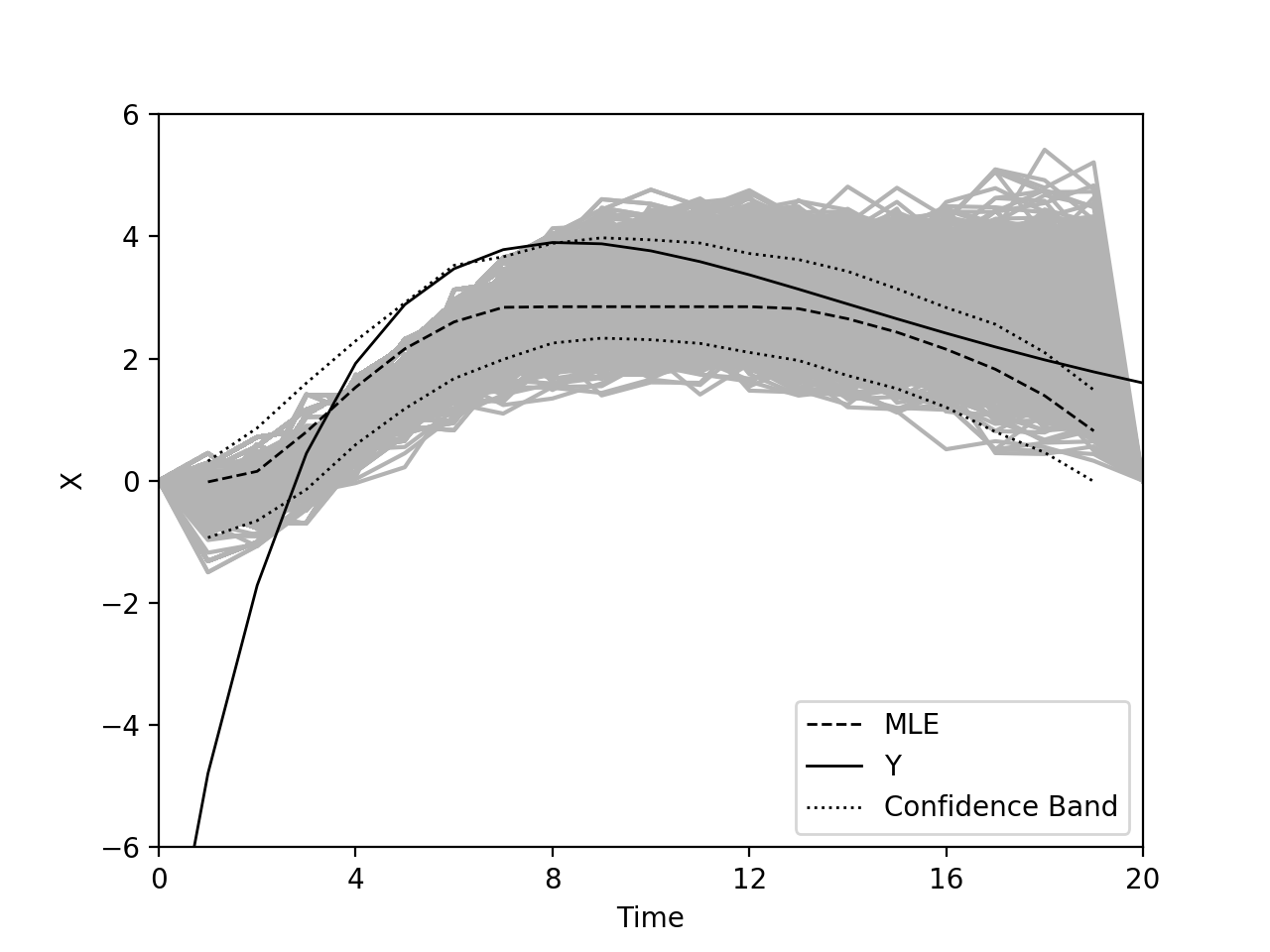}
\includegraphics[width=0.4\textwidth]{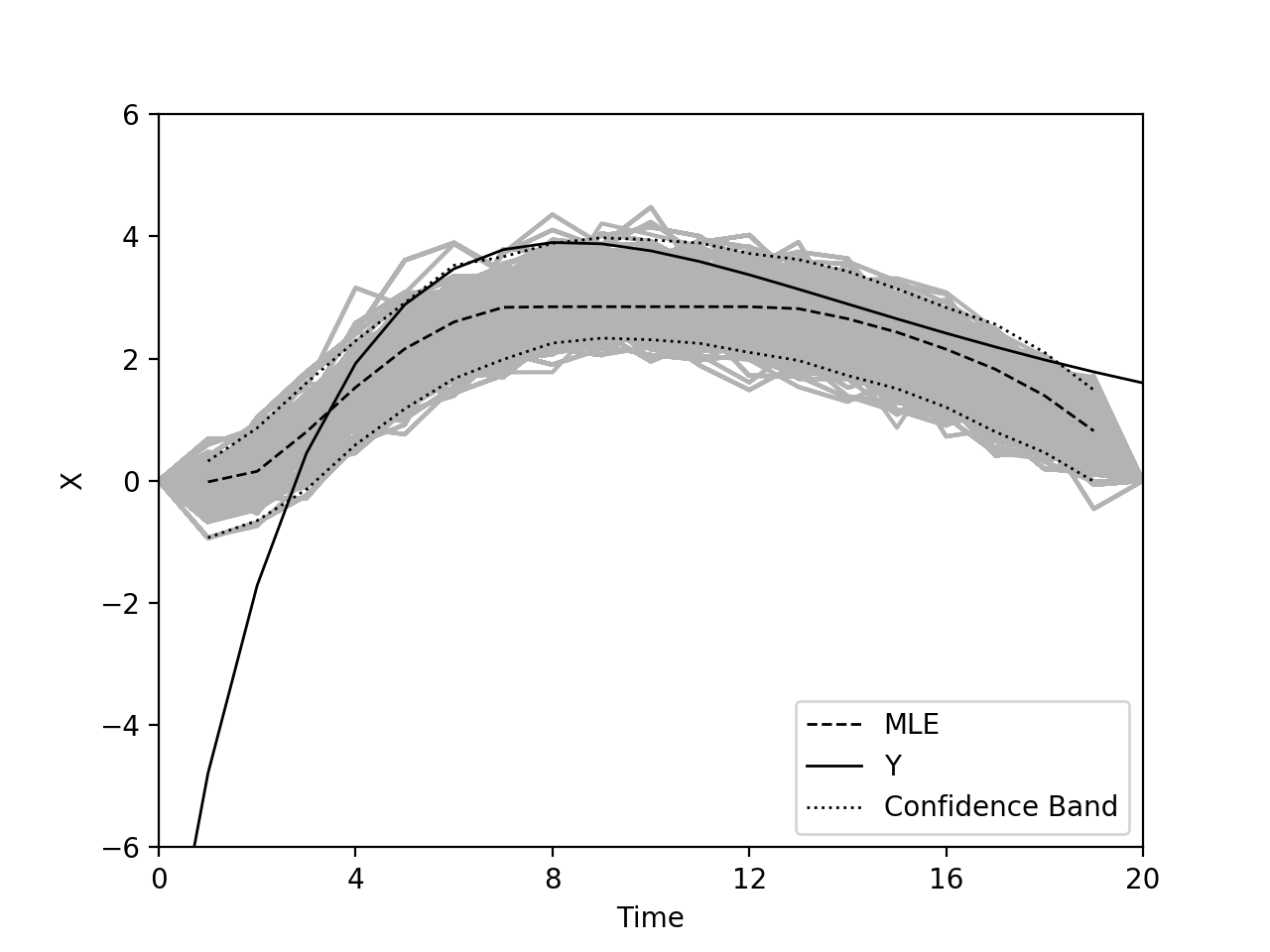}
\caption{Sample paths from the standard SMC method (left panel)
and from the cSMC-BP method (right panel)  before weight adjustment when $\alpha=0.5$. }
\label{fig:cSMC-BP2}
\end{figure}

When $\alpha=0.5$, the state space model is non-Gaussian, hence there is no analytic solution to maximize
the utility function (\ref{eq:optimization}). In this case, we run a
standard SMC sampling with $n=1,000,000$ sample paths to
obtain the most likely sample path, the sample path with the largest likelihood value, together with 95\%  point-wise confidence intervals.
The sample paths generated by SMC and cSMC-BP before weight adjustment, along with
the most likely path and the 95\% confidence region are plotted in
Figure \ref{fig:cSMC-BP2}.
Guided by the priority scores with future information, most samples generated by the cSMC-BP method stay within the 95\% confidence region.

\begin{figure}[hthp]
\centering
\includegraphics[width=0.45\textwidth]{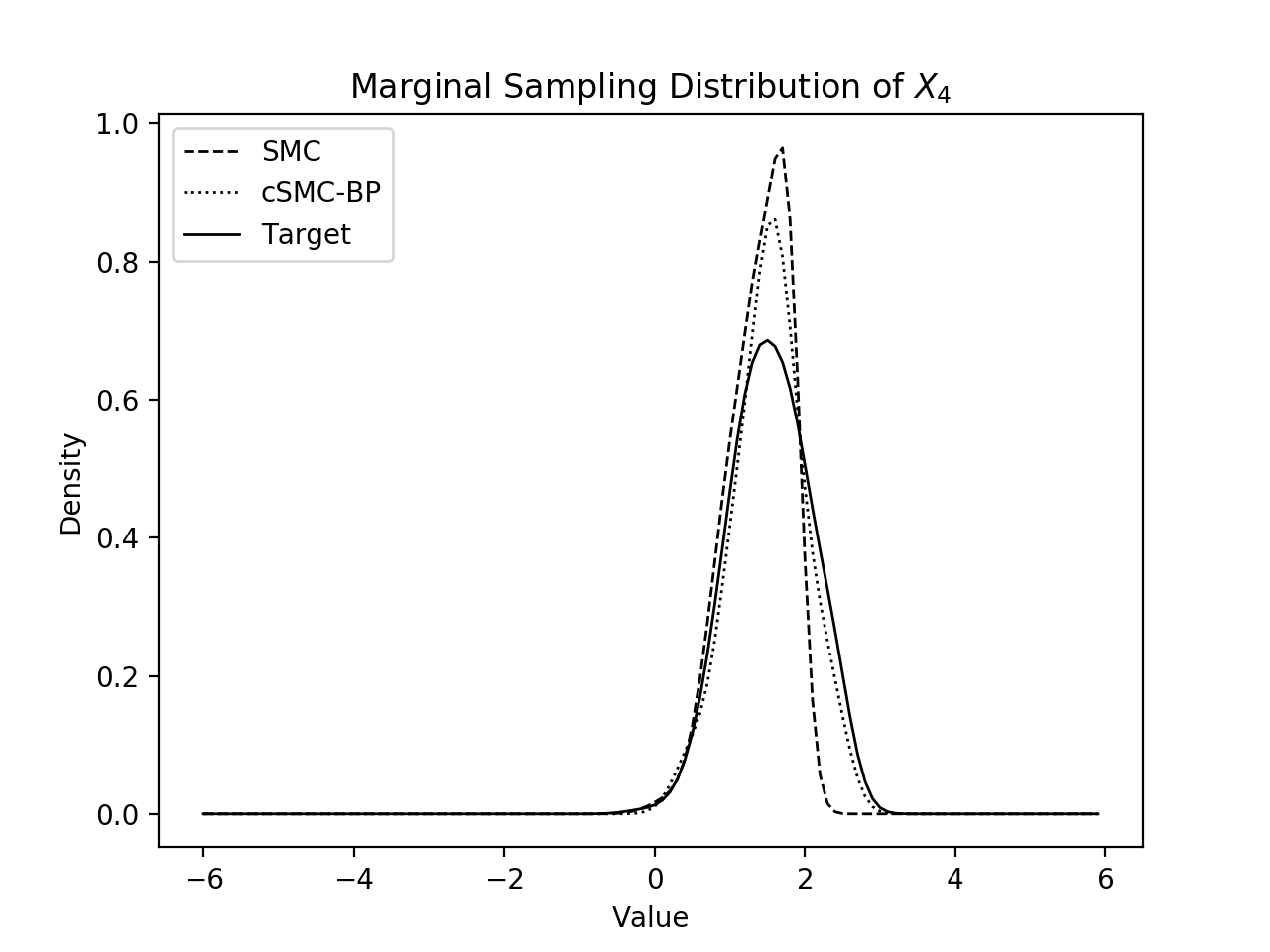}
\includegraphics[width=0.45\textwidth]{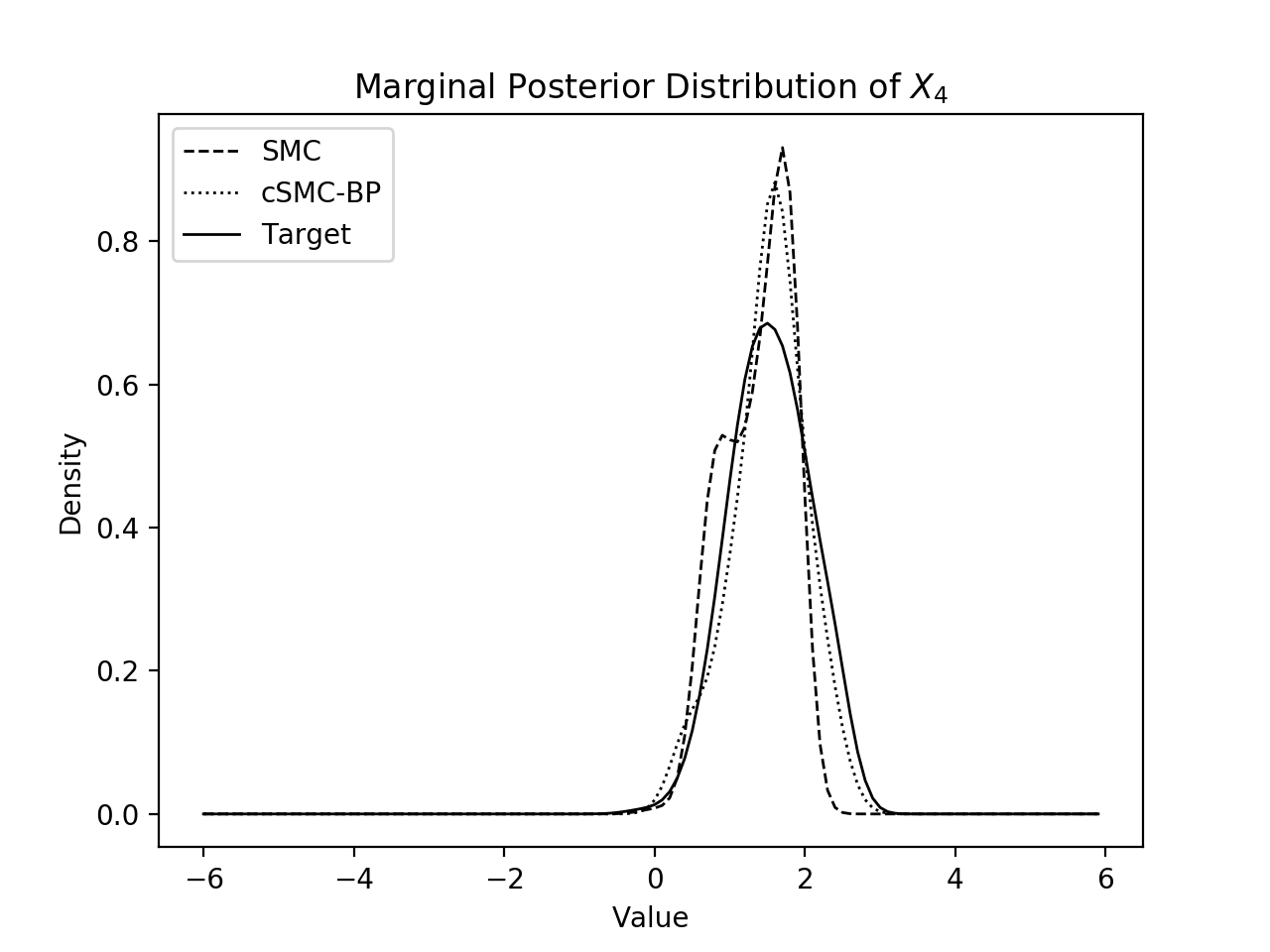}\\
\includegraphics[width=0.45\textwidth]{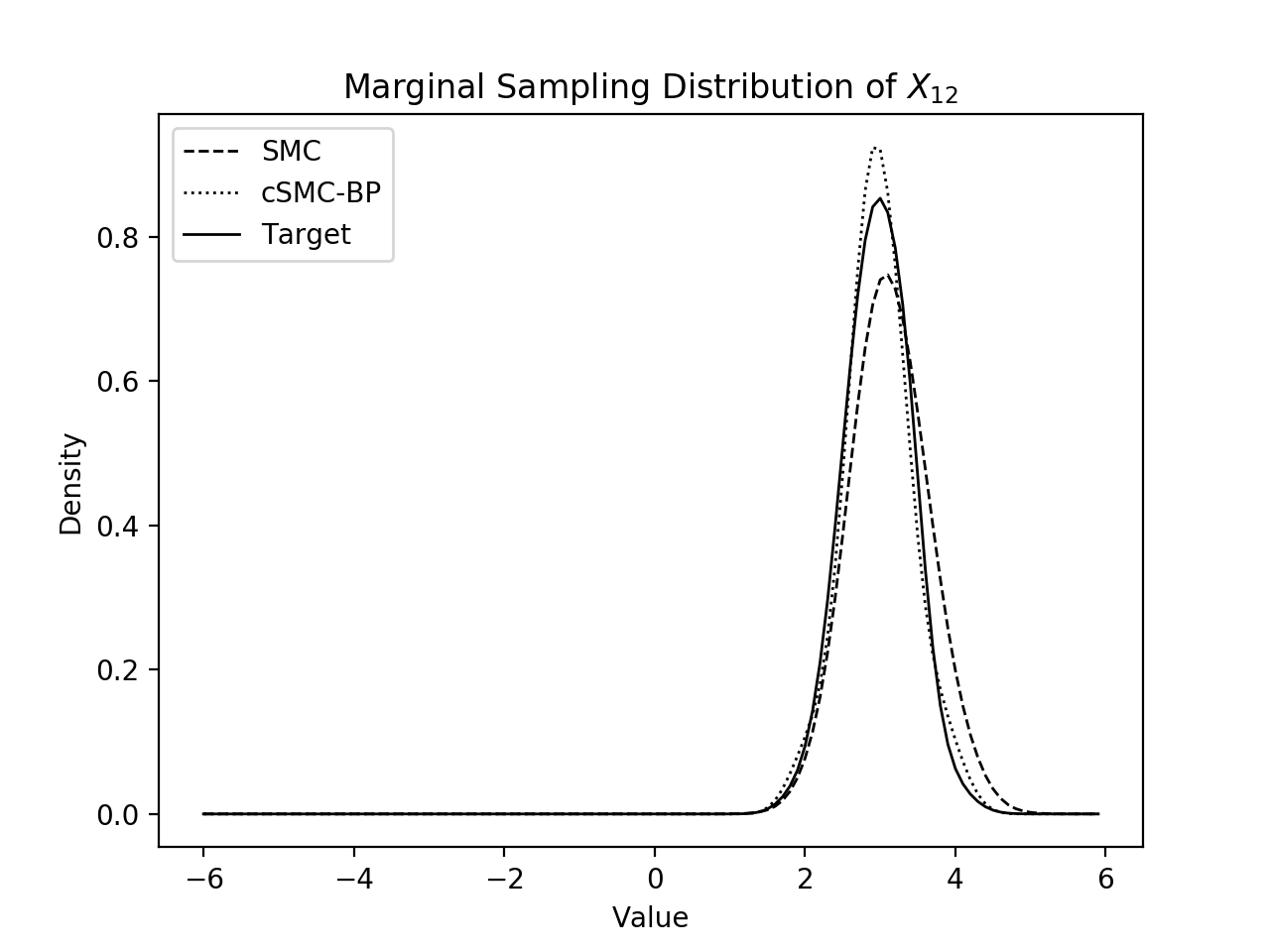}
\includegraphics[width=0.45\textwidth]{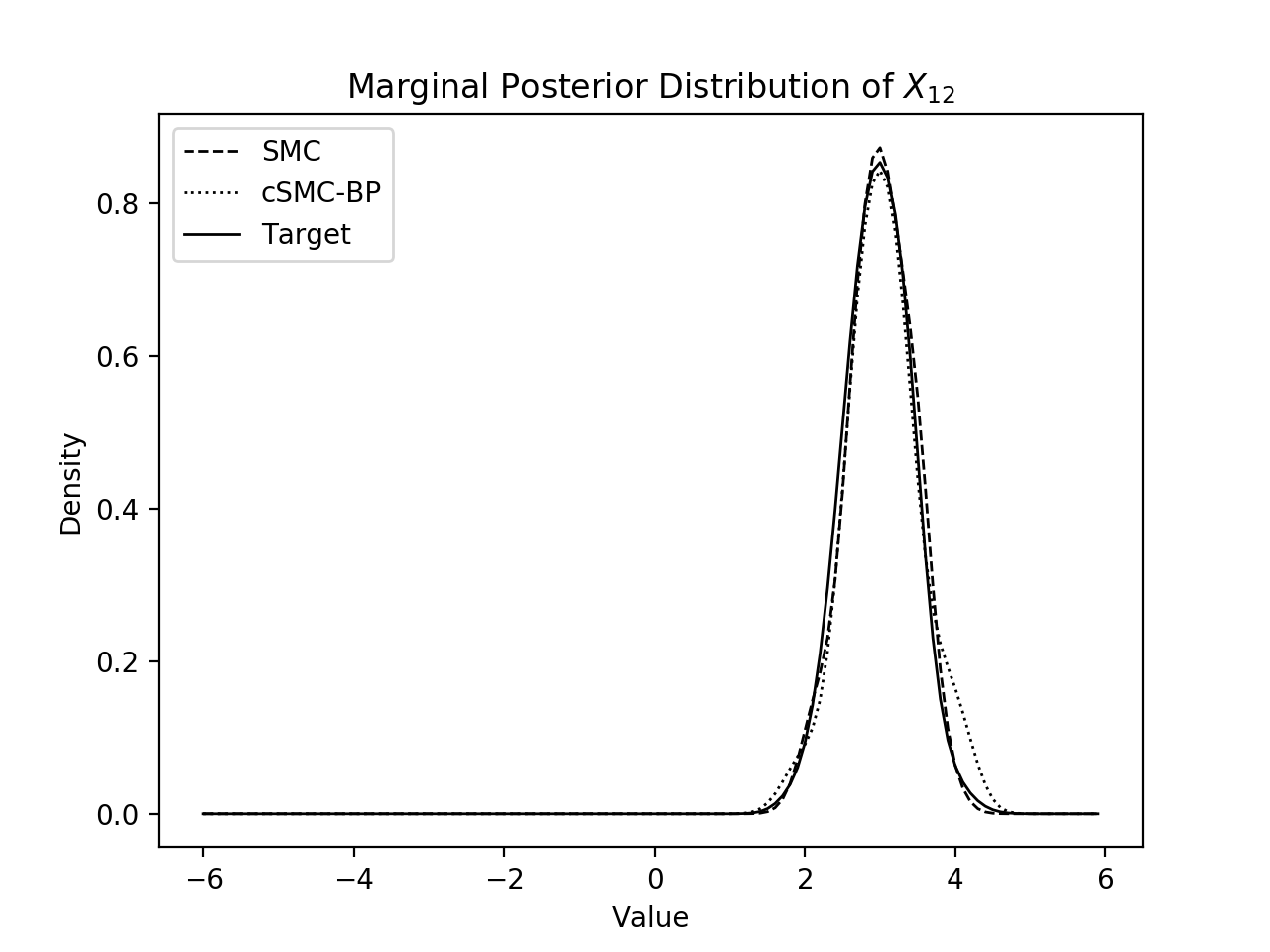}\\
\includegraphics[width=0.45\textwidth]{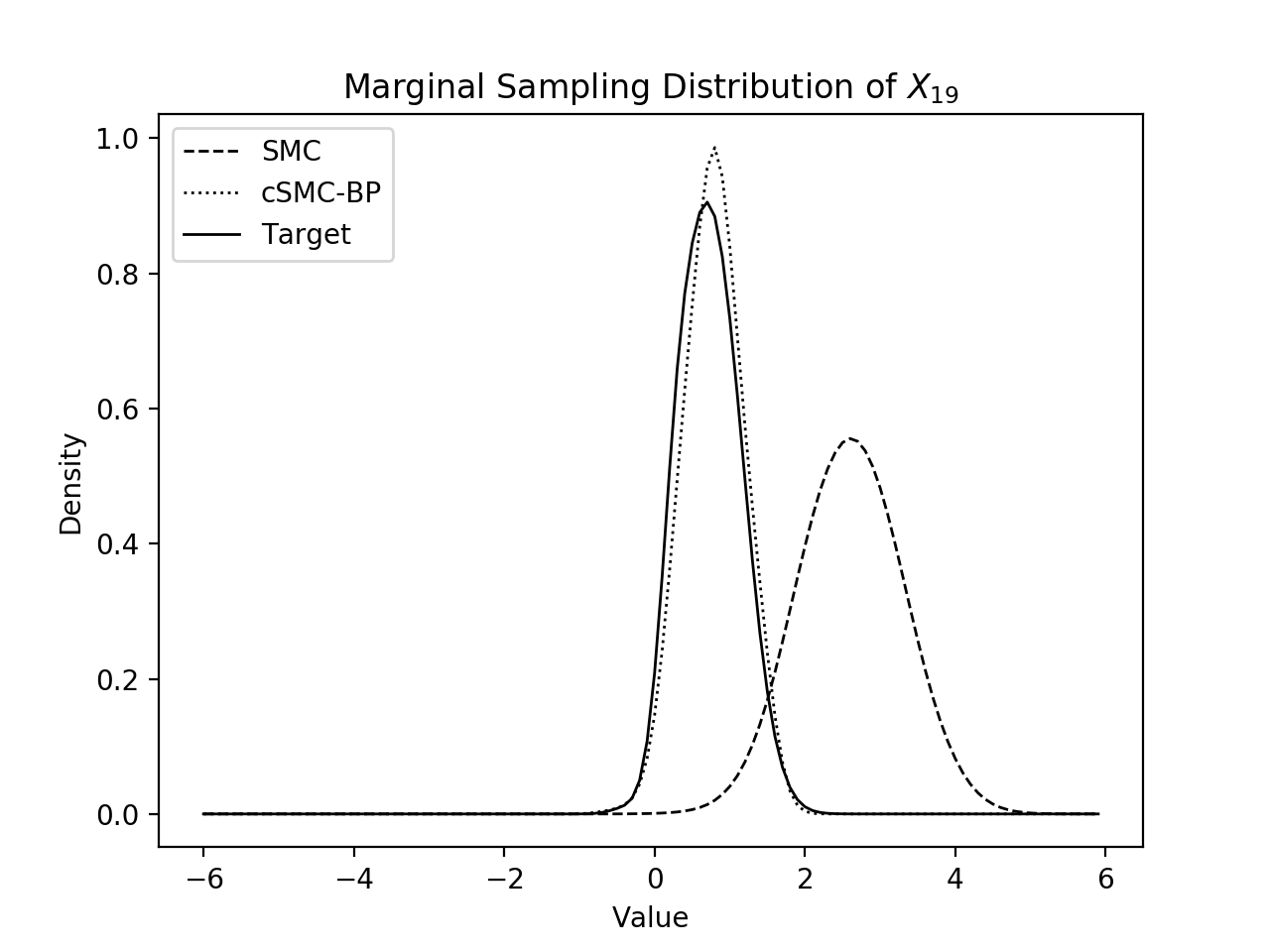}
\includegraphics[width=0.45\textwidth]{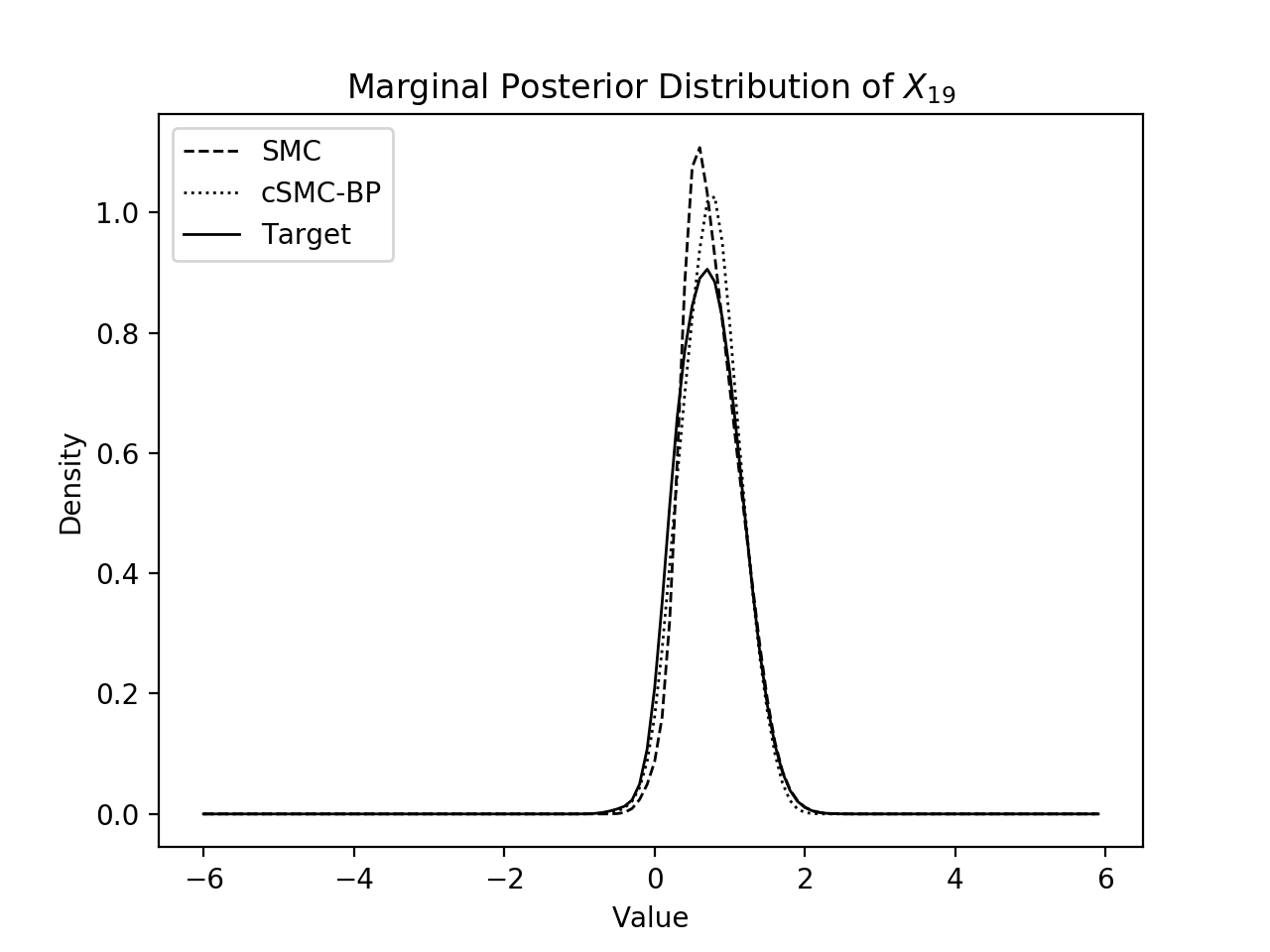}
\vspace{-0.5cm}\caption{Marginal densities of the samples generated by
SMC and cSMC-BP before weight adjustment (left column) and after weight adjustment (right column) at
time $t=4$ (row 1), $t=12$ (row 2)  and $t=19$ (row 3) when $\alpha=0.5$. }
\label{fig:Marginal2}
\end{figure}

\begin{figure}[hthp]
\center
\includegraphics[width = 0.4\textwidth]{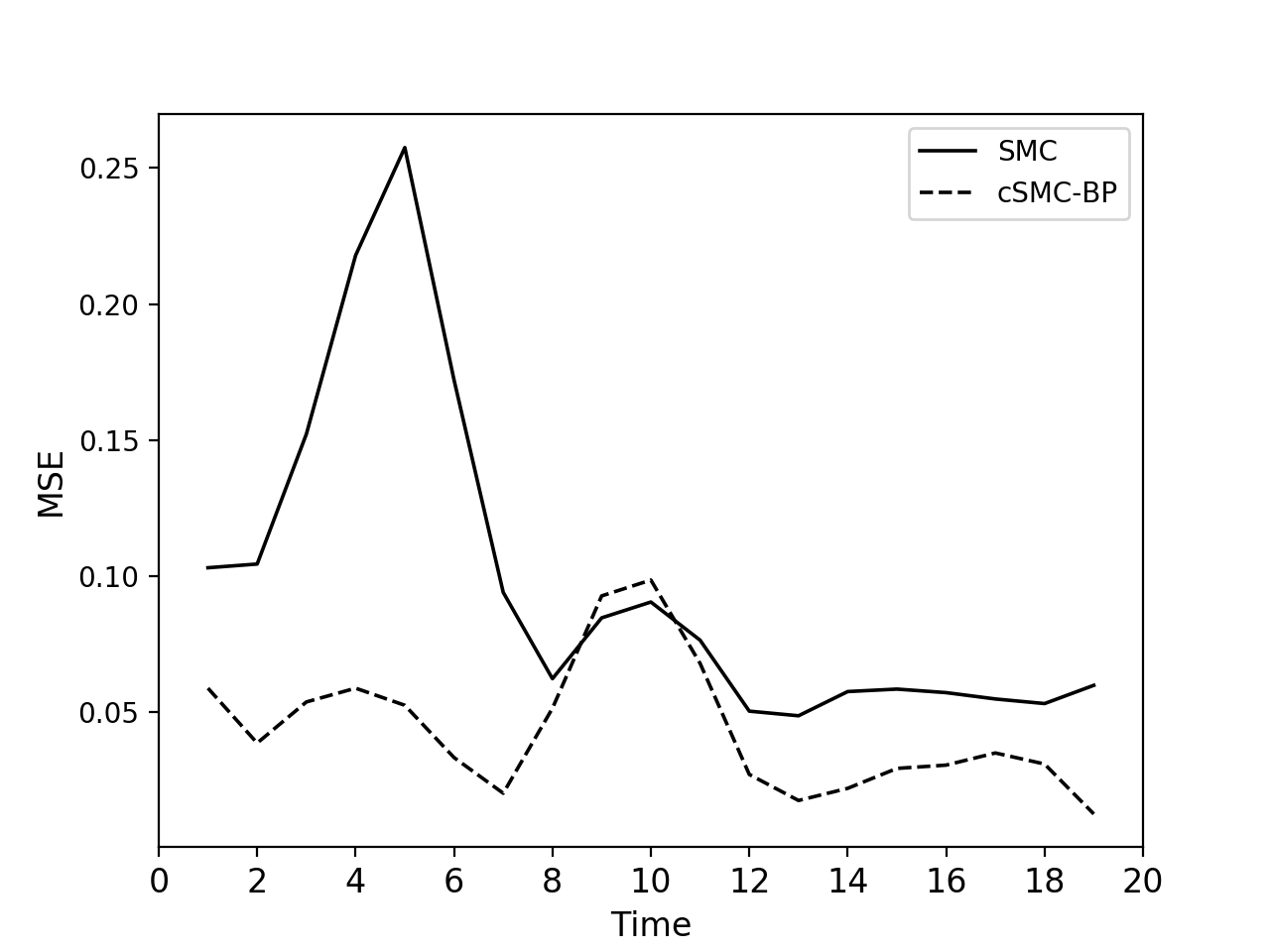}
\caption{ Mean squared error curves for SMC and cSMC-BP when $\alpha=0.5$.}
\label{fig:mse2}
\end{figure}

Figure \ref{fig:Marginal2} plots the marginal densities of the sample paths before weight adjustment (left column) and after weight adjustment (right column).
The true marginal posterior distribution $p(x_t\,|\, x_0,y_{1:T-1},x_T)$ is estimated from the same $n=1,000,000$ SMC sample paths.
At time $t=19$, the distribution of cSMC-BP samples is much closer to the target one than that of SMC samples.
Figure \ref{fig:mse2} plots the MSE's defined in (\ref{eq:MSE}).
The results suggest that cSMC-BP reduces
MSE at most times, especially in the periods $1\leqslant t\leqslant 7$ and $13\leqslant t\leqslant 19$.

\subsubsection{Optimizing the Utility Function}

The Viterbi algorithm  \citep{viterbi1967error, forney1973viterbi} is a dynamic programming algorithm to find the most likely trajectory
in a finite-state hidden Markov model.
In this example, we discretize the state space based on the generated Monte Carlo state
samples to utilize the Viterbi algorithm to
find the optimal path and the optimal value of the utility function $u(x_{0:T})$ in (\ref{eq:optimization}).
Specifically, given $\mathcal X_t=\{x_t^{(i)}\}_{i=1,\cdots,n}$ being the collection of samples of $x_t$ generated by SMC or cSMC-BP, the optimal path
$(\hat x_1, \cdots,\hat x_{T-1})$ is found by solving
the following optimization problem
\ben
(\hat x_1, \cdots,\hat x_{T-1}) = \argmax_{x_1\in\mathcal X_1, \dots, x_{T-1}\in\mathcal X_{T-1}} u(x_{0:T}) 
\een
with the Viterbi algorithm.

\begin{figure}
\centering
\includegraphics[width=0.45\textwidth]{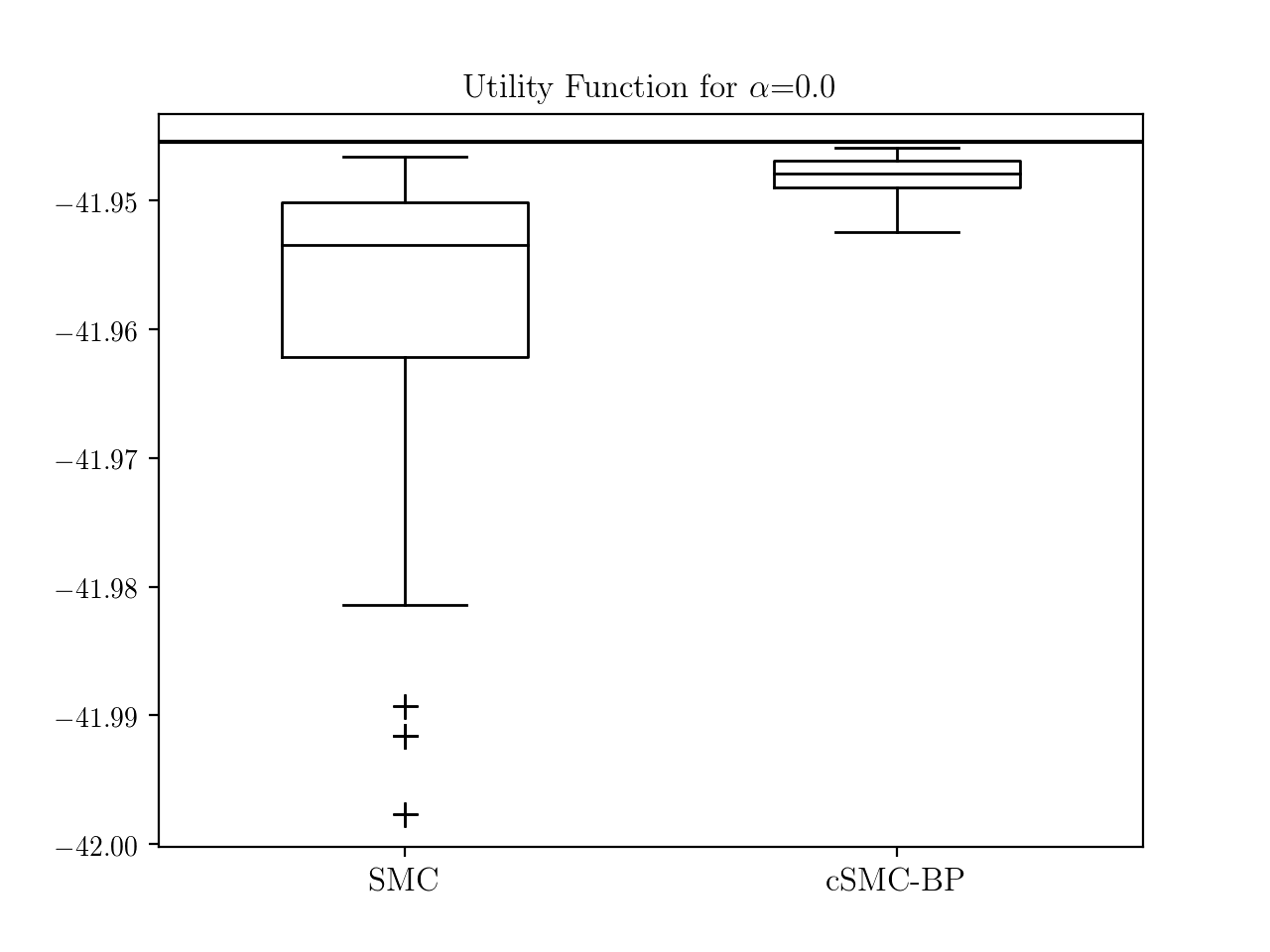}
\includegraphics[width=0.45\textwidth]{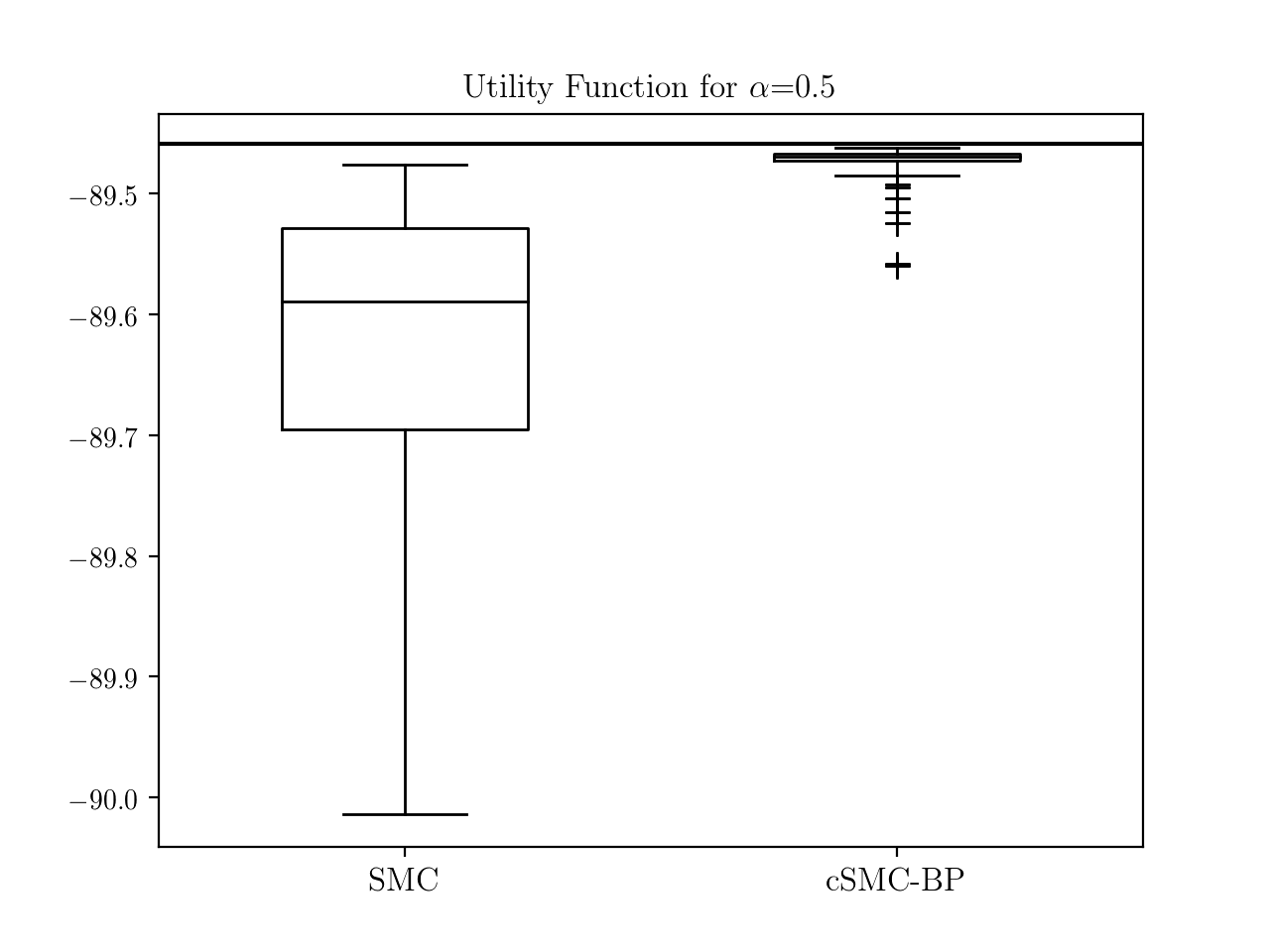}
\caption{Boxplots of optimal values of utility function (\ref{eq:optimization}) solved by the Viterbi algorithm based on
SMC samples and cSMC-BP samples when $\alpha=0$ (left panel)  and $\alpha=0.5$ (right panel). The horizontal lines
are the true optimal values.}\label{fig:viterbi}
\end{figure}

In this experiment,
we use $m=300$ backward pilots and generate $n=500$ Monte Carlo forward samples from cSMC-BP.
For comparison, $n=800$ samples are generated from the standard SMC method.
The experiment is replicated 1,000 times.
The optimal values of the utility function (up to a constant) solved by the Viterbi algorithm based on SMC samples and cSMC-BP samples respectively
are reported in the boxplots in Figure~\ref{fig:viterbi}.
The true optimal value is marked by the horizontal lines. When $\alpha=0.0$, the true optimal value is obtained by the Kalman filter.
When $\alpha=0.5$, the "true" optimal value is computed by the Viterbi algorithm based on a large number ($n=10,000$) of SMC samples.
Compared to the standard SMC method, the cSMC-BP method generates more samples around the true optimal path in the same amount of computation time
by incorporating future information through resampling,
hence it creates a better discrete state space for the Viterbi algorithm. As a result,
the Viterbi algorithm based on cSMC-BP samples can produce trading paths with
larger utility function values for both $\alpha=0$ and $\alpha=0.5$ cases.




\section{Summary}
In this article, we formulate the constrained sampling problem for general stochastic processes and proposed a general framework of cSMC algorithms. The key idea is to use the next available strong information $\mathcal I_{t_+}$ to adjust the sampling distribution at each intermediate point $t$ by choosing appropriate priority scores for resampling. We show that effective priority scores can be obtained from either forward pilot samples or backward pilot samples.

This framework is compatible with previous studies on state space model and diffusion bridge sampling problems as they can be viewed as special cases of the constrained sampling
problems. The sampling procedure of cSMC coincides with the standard SMC approach for a state space model and \cite{lin2010}'s algorithm in the diffusion bridge sampling problem.

Our framework can deal with a wider range of constraints. Three examples are demonstrated in Section 5: one with a subset constraint on the end point,
one with noisy intermediate observation constraints and the other with multilevel constraints. These constraints go beyond the scope of fixed points, but can still be solved using cSMC.

Compared with the standard SMC algorithm, cSMC reduces the divergence between the sampling distribution and the true underlying target distribution by taking future information into consideration at each intermediate step. The additional computational cost is limited. Consequently, cSMC achieves a smaller estimation error with the same amount of computation time
than the standard SMC implementation, as illustrated in the synthetic examples.

\vspace{0.2cm}

\appendix
\bibliography{CSMC_updated}
\bibliographystyle{asa}

\end{document}